\documentclass[twocolumn, prx,showpacs,superscriptaddress]{revtex4-2}
\usepackage{slashbox}
\usepackage{graphicx}
\usepackage{dcolumn}
\usepackage{bm}
\usepackage{xspace}
\usepackage{amsmath}
\usepackage{amsfonts}
\usepackage{amssymb}
\usepackage{xcolor}
\usepackage{float}
\usepackage{multirow}
\usepackage{todonotes}
\usepackage{hyperref}
\usepackage{booktabs}
\usepackage{afterpage}
\usepackage{longtable}

\definecolor{link}{rgb}{0.1,0.1,0.9}
\hypersetup{colorlinks=true,linkcolor=link,citecolor=link,urlcolor=link,linktocpage}
\newcolumntype{C}[1]{>{\PreserveBackslash\centering}p{#1}}

\newcommand{\MNO}{Mn$_{4}$Nb$_{2}$O$_{9}$\xspace}
\newcommand{\MTO}{Mn$_{4}$Ta$_{2}$O$_{9}$\xspace}
\newcommand{\FNO}{Fe$_{4}$Nb$_{2}$O$_{9}$\xspace}
\newcommand{\FTO}{Fe$_{4}$Ta$_{2}$O$_{9}$\xspace}
\newcommand{\CTO}{Co$_{4}$Ta$_{2}$O$_{9}$\xspace}
\newcommand{\CNO}{Co$_{4}$Nb$_{2}$O$_{9}$\xspace}

\newcommand{\sg}{$P\,\bar{3}\,c\,1$}

\newcommand{\ABO}{$\mathcal{A}$$_{4}$$\mathcal{B}$$_{2}$O$_{9}$ ($\mathcal{A}$ = Mn, Fe, Co and $\mathcal{B}$ = Nb, Ta)\xspace}
\newcommand{\ABOshort}{$\mathcal{A}$$_{4}$$\mathcal{B}$$_{2}$O$_{9}$\xspace}

\newcommand{\FNOx}{FeNb$_{2}$O$_{6}$\xspace}

\newcommand{\CrO}{Cr$_{2}$O$_{3}$\xspace}

\begin{document}
\preprint{APS/123-QED}

\title{High-resolution neutron diffraction determination of noncollinear antiferromagnetic order in the honeycomb magnetoelectric \FNO}

\author{Raktim Datta}
\affiliation{Center for Van der Waals Quantum Solids, Institute for Basic Science (IBS), Pohang 37673, Republic of Korea}

\author{Kapil Kumar}
\affiliation{Center for Van der Waals Quantum Solids, Institute for Basic Science (IBS), Pohang 37673, Republic of Korea}

\author{Dong Gun Oh}
\affiliation{Department of Physics, Yonsei University, Seoul 03722, Republic of Korea}

\author{Dongwook Kim}
\affiliation{Department of Physics, Chonnam National University, Gwangju 61186, Republic of Korea}

\author{Rahul Goel}
\affiliation{Center for Van der Waals Quantum Solids, Institute for Basic Science (IBS), Pohang 37673, Republic of Korea}
\affiliation{Center for Integrated Nanostructure Physics, Institute for Basic Science (IBS), Suwon 16419, Republic of Korea}
\affiliation{Sungkyunkwan University, Suwon 16419, Republic of Korea}

\author{Nara Lee}
\affiliation{Department of Physics, Yonsei University, Seoul 03722, Republic of Korea}

\author{Ara Go}
\affiliation{Department of Physics, Chonnam National University, Gwangju 61186, Republic of Korea}

\author{Young Jai Choi}
\affiliation{Department of Physics, Yonsei University, Seoul 03722, Republic of Korea}

\author{Valery Kiryukhin}
\affiliation{Department of Physics and Astronomy, Rutgers University, Piscataway, New Jersey 08854, USA}

\author{Sungkyun Choi}
\email{sungkyunchoi@ibs.re.kr}
\affiliation{Center for Van der Waals Quantum Solids, Institute for Basic Science (IBS), Pohang 37673, Republic of Korea}
\affiliation{Center for Integrated Nanostructure Physics, Institute for Basic Science (IBS), Suwon 16419, Republic of Korea}
\affiliation{Sungkyunkwan University, Suwon 16419, Republic of Korea}
\affiliation{Department of Physics and Astronomy, Rutgers University, Piscataway, New Jersey 08854, USA}

\date{\today}

\begin{abstract}
Magnetoelectric systems offer potential for device applications exploiting coupled states between electric and magnetic properties. Among magnetoelectric materials, \FNO has attracted special attention because of its pronounced dielectric signal at high magnetic transition temperatures. However, the magnetic ground state, which is essential information for understanding its unusual magnetoelectricity, remains unclarified. Here, we report a noncollinear magnetic ground state of \FNO. To examine the magnetoelectric effect associated with sequential magnetic and structural transitions upon cooling, we conducted combined x-ray diffraction, magnetic susceptibility, magnetization, dielectric constant, and magnetodielectric experiments. Powder neutron diffraction experiments revealed a series of magnetic Bragg peaks and clear splitting of peaks via structural transition. Magnetic Rietveld refinements, combined with group theory analysis, determined a noncollinear antiferromagnetic structure including a significant $c$-axis moment component at 1.5~K. This study provides insights into the understanding of its magnetoelectric properties.
\end{abstract}

\maketitle

\section{Introduction}
\label{sec:intro}
Magnetoelectric (ME) and multiferroic materials are compounds that exhibit coupling between the electric and magnetic orders~\cite{Cheong2007, Spaldin2019}. This coupling enables magnetic (electric polarization) control via electric (magnetic) fields. Such coupling holds fundamental significance and promises technological applications, including energy-efficient and next-generation memory devices~\cite{Eerenstein2006}.

The study of magnetoelectrics was triggered by an original theoretical conjecture on the ME effect by considering symmetries~\cite{Dzyaloshinskii1960}, where spatial-inversion and time-reversal symmetry breaking enabled the coupled state and was subsequently verified experimentally in Cr$_2$O$_3$~\cite{Astrov1960}. Its magnetic state lacks ferroelectricity under ambient conditions but responds proportionally to an external magnetic field through inversion symmetry breaking---a characteristic termed ``linear magnetoelectricity" for this material class.

In type-II multiferroics, magnetically induced ferroelectricity arises because the magnetic ordering breaks the inversion symmetry, generating electric polarization~\cite{Tokura2014}. Well-known examples include TbMnO$_{3}$~\cite{Kimura2003} and TbMn$_{2}$O$_{5}$~\cite{Hur2004}. However, transition temperatures are typically limited to a few tens of kelvins, hampering practical applications. Therefore, materials with strong ME or unconventional effects are urgently needed for technological applications.

\begin{figure}[t]
\includegraphics[width=\linewidth]{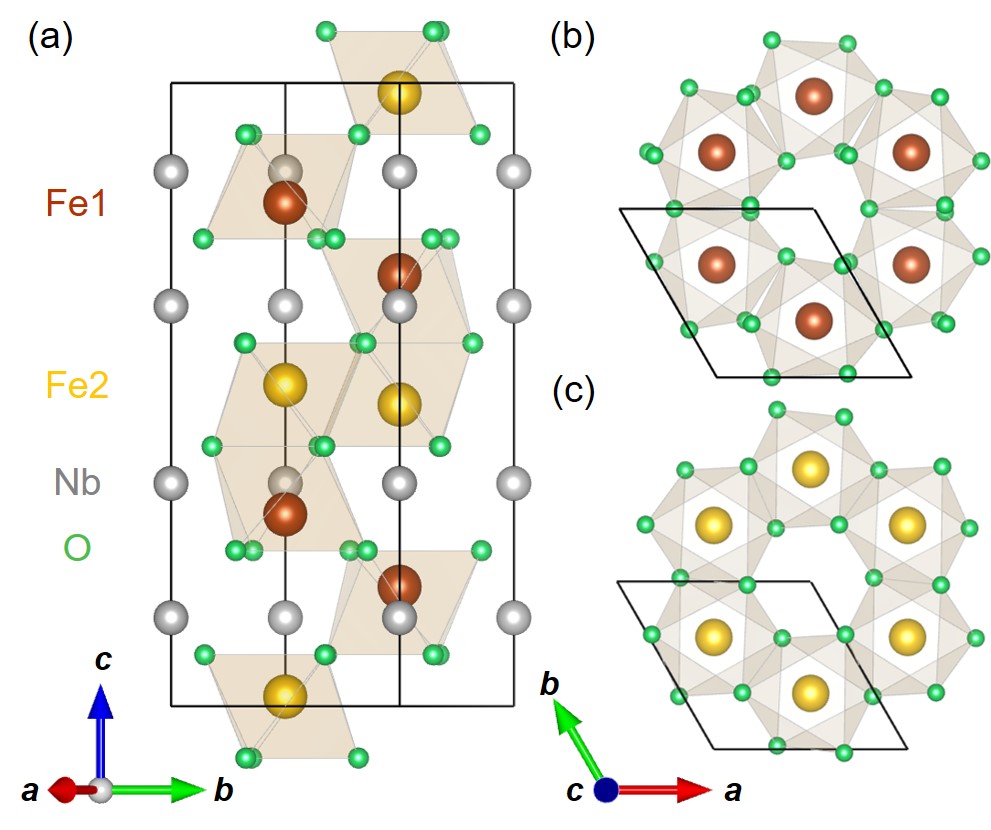}
	\caption{Crystal structure of \FNO. (a) The crystallographically distinct Fe1 and Fe2 sites in a unit cell (black solid lines, 0~$\leqslant$~\textit{z}~$\leqslant$~1). (b) and (c) The buckled (0.15~$\leqslant$~$z$~$\leqslant$~0.35) and flat ($-$0.1~$\leqslant$~$z$~$\leqslant$~0.1) honeycomb layers.
}
	\label{fig:str}
\end{figure}

Honeycomb-based ME antiferromagnets \ABO~\cite{Bertaut1961} have recently attracted significant interest. They form a three-dimensional lattice in a trigonal unit cell (Fig.~\ref{fig:str}) and deviate from the well-known corundum crystal structure observed in \CrO~\cite{Dzyaloshinskii1960}, comprising an alternating stacking of two different honeycomb layers along the \textit{c} axis [see Figs.~\ref{fig:str} (b) and \ref{fig:str}(c)]. In \ABOshort, the interplay between different valence states of the magnetic ions and two sublattices results in versatile ME properties, such as magnetodielectric effects in \MNO~\cite{Fang2015}, anisotropic magnetoelectricity in \MTO~\cite{Panja2021}, manipulation of electric polarization with a rotating magnetic field within the honeycomb plane in \CNO~\cite{Khanh2017}, and nonlinear and anisotropic ME effects in \CTO~\cite{Lee2020}. Fe-based compounds have also exhibited rare coexistence of linear ME and type-II multiferroism in \FTO~\cite{Maignan2018:2}, as well as nonlinear ME effects and electric control of magnetism in \FNO~\cite{Zhang2023}.

In this study, \FNO is examined because of its unusual ME properties. Its magnetic transition temperature is relatively high ($\sim$90~K)~\cite{Maignan2018} with an exceptionally large magnetic-order-induced dielectric change~\cite{Ding2020}. Furthermore, \FNO exhibits the largest polarization within the family~\cite{Zhang2023}. However, the origins of these effects are unclear because of an ongoing controversy regarding the magnetic order and magnetic space group (MSG). The antiferromagnetic structure of \FNO was initially proposed to be collinear in the $ab$ plane by powder neutron diffraction~\cite{Jana2019} in $C2/c'$ (No.~15.88) and $C2'/c$ (No.~15.87) MSGs. The noncollinear antiferromagnetic order within the plane was later found by single-crystal neutron diffraction~\cite{Ding2020} in the $C2'/c$ MSG. However, this cannot explain all nine components in the ME tensor experimentally observed below $T_{\rm N}$~\cite{Zhang2023} because the $C2'/c$ MSG allows only four components~\cite{Choi2020}, suggesting that the $P\bar{1}'$ (No.~2.6) MSG could be the ground state, which is the highest possible symmetry. As a result, the magnetic structure of \FNO remains under debate.

We determined the noncollinear magnetic order of \FNO including a finite magnetic moment along the $c$ axis, via high-quality neutron diffraction. The polycrystalline powder was carefully grown by minimizing a magnetic secondary impurity phase, which was confirmed by x-ray diffraction analysis, followed by examination of the magnetic and dielectric properties. Subsequent temperature-dependent neutron diffraction measurements were performed, and the data were analyzed via magnetic Rietveld refinements complemented with group theory. A more accurate magnetic structure was obtained, including a magnetic moment along the $c$ axis, in contrast to two previously reported findings~\cite{Jana2019, Ding2020}. We also performed a systematic analysis using lower magnetic space groups. Our results provide essential information about the magnetic structure, MSG, and therefore the magnetic point group, giving insights into understanding the complex ME effect of \FNO.

The remainder of this paper is organized as follows. Sections~\ref{sec:exp} and \ref{sec:comp} describe the details for experiments and calculations, respectively. Section~\ref{sec:res} presents the results: ME properties (Sec.~\ref{sec:ME:properties}), successive magnetic and structural transitions observed in neutron diffraction (Sec.~\ref{sec:ND:TD}), and neutron diffraction with group theory analysis (Sec.~\ref{sec:ND:mag:str}). Section~\ref{sec:discussion} discusses the implications, and Sec.~\ref{sec:conclusions} concludes the study. The Appendix includes additional information.

\section{Experimental details}
\label{sec:exp}

\begin{figure*}[t]
\includegraphics[width=\linewidth]{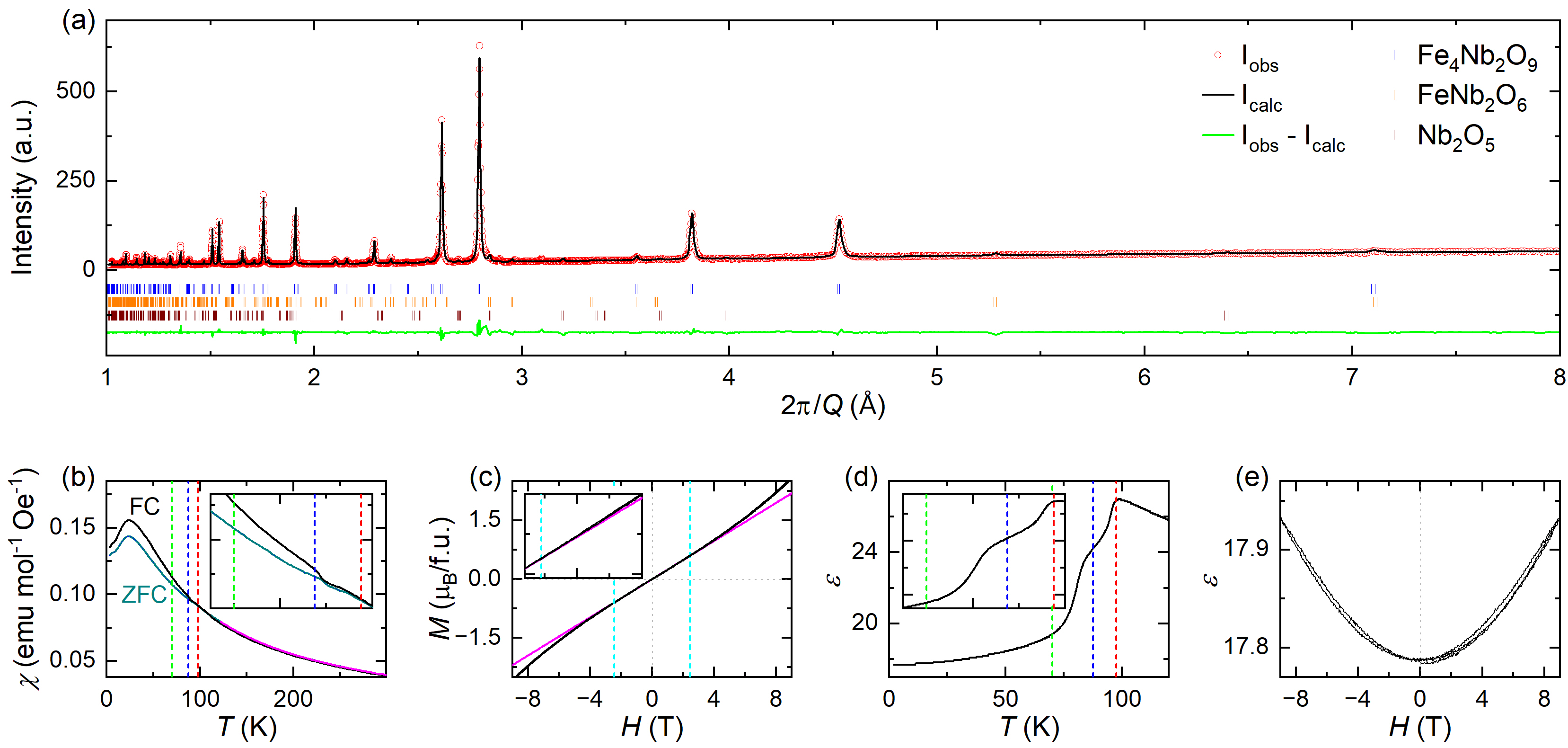}
\caption{Sample characterizations. (a) Rietveld refinement results obtained from x-ray diffraction patterns collected at room temperature. Impurity peaks are minimal. (b) ZFC and FC magnetic susceptibilities at 0.05~T. The solid magenta line indicates the CW fit (see text). (c) Magnetic isotherm at 2.5 K. The solid magenta line is a linear fit. The dashed cyan lines denote the critical magnetic field for the spin-flop transition. (d) Dielectric constant at 0.05~T. (e) Magnetodielectric data for the magnetic field at 2~K. The dashed red, blue, and green lines are the $T_{\rm N}$, $T_{\rm S}$, and $T_{\rm 0}$ values determined from the neutron diffraction data (Fig.~\ref{fig:Block 3}).
}
\label{fig:Char}
\end{figure*}

Polycrystalline \FNO samples were prepared by a solid-state reaction, as described elsewhere~\cite{Maignan2018}. Stoichiometric amounts of Fe, Fe$_{2}$O$_{3}$, and Nb$_{2}$O$_{5}$ powders were thoroughly mixed and then pelletized in a glove box ﬁlled with argon. The pellets were then sealed under a vacuum in a quartz tube and heated at 1000$^{\circ}$C for 15 h in a box furnace. The calcined pellet was finely reground and sintered at 1100$^{\circ}$C for 24 h. Powder X-ray diffraction was conducted at room temperature to verify the main phase and minimize the magnetic impurity phase \FNO~\cite{Heid1996, Ding2020} using Cu \textit{K}$\alpha$ radiation. Magnetic susceptibility was measured at 0.05~T between 2~and 300~K, and magnetization data were collected at 2.5~K (from $-$9~to 9~T) using a vibrating sample magnetometer in a physical properties measurement system (Quantum Design). The dielectric constant was measured at 100~kHz using an LCR meter for magnetic field (from $-$9~to 9~T) and temperature.

Time-of-flight neutron diffraction measurements were performed on the optimally grown polycrystalline sample on Wide angle in a Single Histogram (WISH) beamline at ISIS, United Kingdom~\cite{Chapon2011}. During the data reduction, time-of-flight data were summed by pairing detector bank data from opposite sides of the scattering angle. The resulting data were categorized into detectors labeled banks 1 to 5 by pairing banks with the same scattering angle on the opposite side of the instrument, corresponding to the ascending order of the specific scattering angles 27.08$^{\circ}$, 58.33$^{\circ}$, 90$^{\circ}$, 121.66$^{\circ}$, and 152.83$^{\circ}$, respectively. Group theory analysis based on the Bilbao crystallographic server identified candidate MSGs~\cite{Aroyo2011}. Nuclear and magnetic refinements were conducted using the JANA2006 package~\cite{Petricek2014}. Reciprocal lattice units are presented in the parent trigonal lattice otherwise specified. Peak indexing for the monoclinic or triclinic lattices is also addressed where applicable.

\section{Computational details}
\label{sec:comp}
First-principles calculations based on density functional theory were performed using the Vienna Ab initio Simulation Package (VASP)~\cite{Kresse1996}. The exchange-correlation energy was treated within the framework of the generalized gradient approximation (GGA), using the Perdew–Burke–Ernzerhof functional~\cite{Perdew1996}. The plane-wave energy cutoff was set to 1100 eV, and the Brillouin zone was sampled using a 3 $\times$ 6 $\times$ 2 Monkhorst-Pack \textit{k}-point mesh~\cite{Monkhorst1976}. The interaction between valence and core electrons was described using the projector augmented-wave method~\cite{Blochl1994, Kresse1999}. To account for the on-site Coulomb interaction of localized Fe 3\textit{d} electrons, the GGA+$U$ formalism was adopted in the Dudarev approach~\cite{Dudarev1998}, with the effective Hubbard \textit{U} parameter ($U_{\rm eff}$ = $U$ -- $J$) varied from 3 to 7 eV.

\section{Results}
\label{sec:res}

\subsection{Coupled magnetic and electric properties}
\label{sec:ME:properties}
This study utilized high-quality polycrystalline \FNO powder with a minimized impurity phase, as confirmed by x-ray diffraction. Figure~\ref{fig:Char}(a) presents Rietveld refinement results on x-ray diffraction patterns obtained at room temperature. The trigonal structure was confirmed (\sg, No.~165) to be consistent with the literature~\cite{Maignan2018, Jana2019, Ding2020}. The Bragg peaks were indexed with the target phase \FNO with tiny impurity peaks from \FNOx, and Nb$_{2}$O$_{5}$ (a level of 1\% mass volume).

\begin{figure*}[t!]
\includegraphics[width=\linewidth]{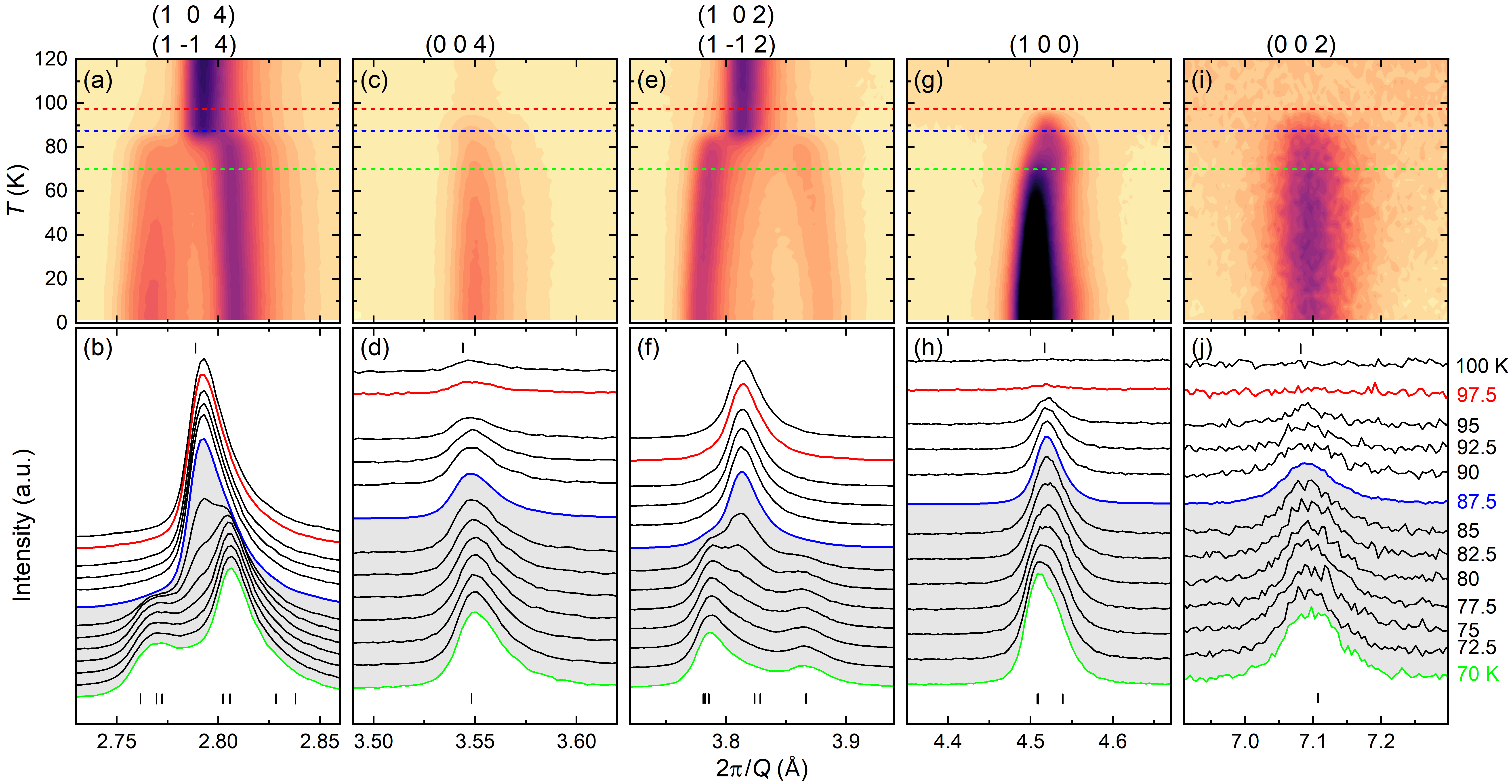}
\caption{Evolution of magnetic and nuclear Bragg peaks upon cooling. (a), (c), (e), (g), and (i) Contour plots for representative magnetic and nuclear Bragg peaks. (b), (d), (f), (g) and (h), and (j) Corresponding diffraction patterns from 100 to 70~K, highlighting two transitions. Red, blue, and green lines represent $T_{\rm N}$, $T_{\rm S}$, and $T_{\rm 0}$. The $T_{\rm 0}$ is the temperature where the low-temperature peaks are only observed. Data from bank 3 are displayed. Peak indexing above the top panel is based on the parent trigonal lattice. Top vertical ticks for peak indexing in (b), (d), (f), (h), and (j) correspond to 100~K in the trigonal lattice, while those in the lower section are from 70~K in the triclinic lattice.
}
\label{fig:Block 3}
\end{figure*}

\begin{figure}[t!]
\includegraphics[width=\linewidth]{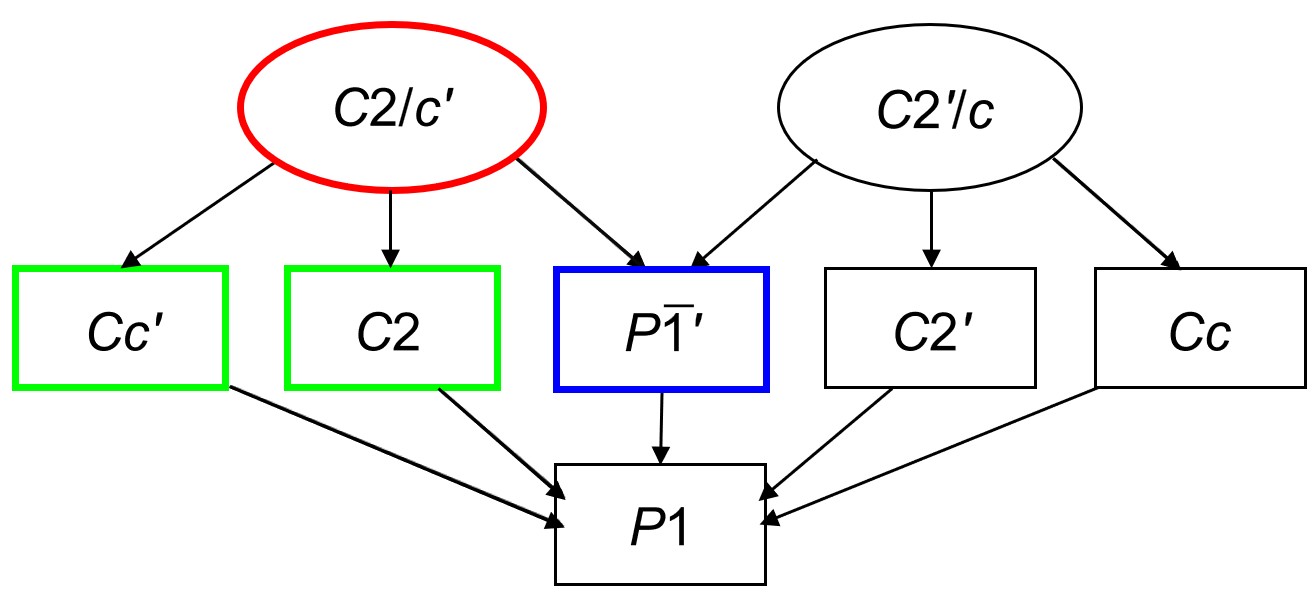}
\caption{Group-subgroup diagram. A hierarchy of MSGs from $C2/c'$ and $C2'/c$ to the lowest possible group, $P1$, is shown. Three potential MSGs are highlighted by the green and blue rectangles.
}
\label{fig:Group_subgroup}
\end{figure}

We then performed a combined measurement of magnetic susceptibility, magnetization, and dielectric constant measurements to examine the magnetoelectric properties of \FNO. Figure~\ref{fig:Char}(b) displays the zero-field-cooled (ZFC) and field-cooled (FC) susceptibilities $\chi$ at 0.05~T. A ZFC-FC bifurcation revealed susceptibility anomaly, indicating a long-range antiferromagnetic transition at $T\rm_{N}$~$\sim$~90 K. This is consistent with neutron diffraction results (which will be discussed later). High-temperature fitting was performed on the ZFC data between 120 and 300~K by using the Curie-Weiss (CW) equation $\chi(T) = \chi_{0} + C/(T-\theta)$, where $\chi_{0}$ is temperature-independent susceptibility, $C$ is the Curie constant, and $\theta$ is the CW temperature~\cite{Ali2024}. We obtained $\mu_\textrm{eff}$~$\sim$~5.17~$\mu_\textrm{B}$/Fe, which is close to $\mu_\textrm{eff}$~$\sim$~4.9~$\mu_\textrm{B}$ for high-spin Fe$^{2+}$ (3$d^{6}$)~\cite{Maignan2018}. A negative $\theta$ ($\sim$~$-$48 K) indicates a dominant antiferromagnetic exchange interaction~\cite{Maignan2018,Chen2021}. Figure~\ref{fig:Char}(c) shows the magnetic isotherm at 2.5~K. It displays a linear feature, suggesting a consistent dominant antiferromagnetic interaction. A spin-flop transition was also observed at around 2.45~T, which was consistently reported in the literature~\cite{Maignan2018,Chaudhary2020,Chen2021}.

The magnetodielectric properties were also measured. Figure~\ref{fig:Char}(d) presents dielectric constant $\varepsilon$ data at 100~kHz at 0.05~T. Upon cooling, $\varepsilon$ increases because the electric field polarizes the material more effectively with the decreased thermal agitation. A rather broad peak is then found at $T\rm_{N}$, possibly due to spin-phonon coupling~\cite{Singh2023}. It also aligns with a magnetic susceptibility anomaly [dashed red line in Fig.~\ref{fig:Char}(b)]. At around $T_{\rm S}$~$\sim$~87.5~K, another peak was observed [marked by the blue dashed line in Fig.~\ref{fig:Char}(b)]. This is a signature structural transition, consistent with previous reports~\cite{Maignan2018, Jana2019}. $\varepsilon$ decreased monotonically and saturated toward the base temperature, indicating no dramatic change in the magnetic and nuclear structures. Therefore, we observed two anomalies in dielectric constant measurements which are consistent with those observed in magnetic susceptibility, as shown in Fig.~\ref{fig:Char}(b). This strongly suggests coupled magnetic and electric properties. We also observed weak nonlinear dielectric properties at 2~K at a magnetic field beyond 3~T [Fig.~\ref{fig:Char}(e)]. This is compatible with a nonlinear ME effect at high magnetic fields, possibly due to spin rotation and magnetic domain realignment~\cite{Zhang2023}.

\begin{figure*}[t!]
\includegraphics[width=\linewidth]{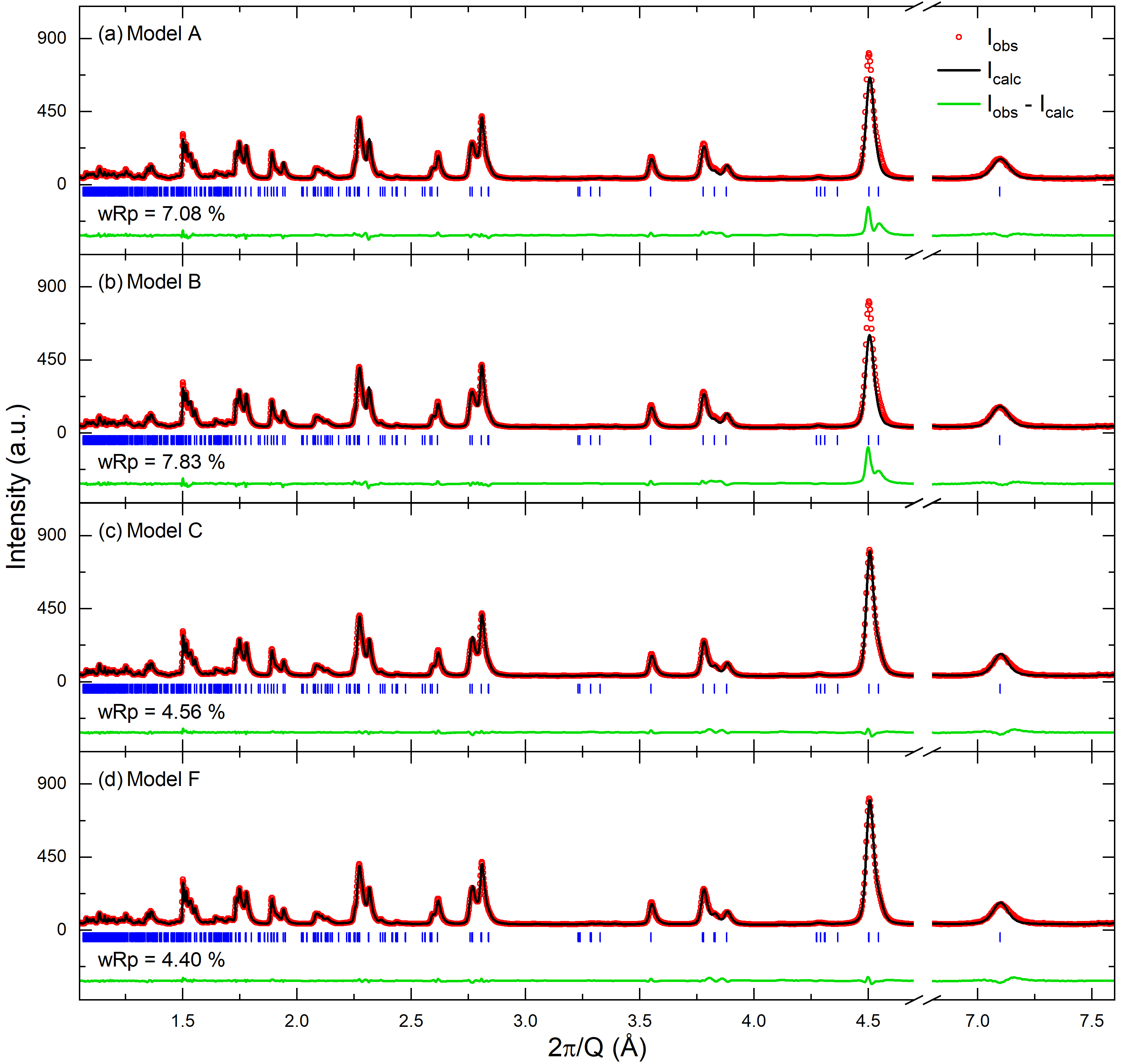}
\caption{Magnetic and nuclear refinements at 1.5~K. (a) Model A: collinear antiferromagnetic order without $M_{c}$~\cite{Jana2019}. (b) Model B: noncollinear antiferromagnetic order without $M_{c}$~\cite{Ding2020}. (c) Model C: noncollinear antiferromagnetic order with $M_{c}$ using $C2/c'$. (d) Model F: noncollinear antiferromagnetic order with $M_{c}$ using $P\bar{1}'$. Bank3 data are presented. The \textit{w}\textit{R}$_{\mathrm{p}}$ values shown are for all banks (see Tables~\ref{1p5K_magnetic_results_right_part} and~\ref{table:1p5K:magnetic:results} for detailed information). Bragg peaks for impurity phases are omitted for brevity and to facilitate clearer matching of the main peaks
	with the data.
}
\label{fig:mag:refine:1p5K}
\end{figure*}

\begin{table} [t]
	\caption{Reliability factors for magnetic structural candidates. B denotes bank. GOF denotes goodness of fit.
	}
	\vspace{0.2cm}
	\label{1p5K_magnetic_results_right_part}
	\setlength\extrarowheight{4pt}
	\setlength{\tabcolsep}{1.5pt}
	\begin{tabular}{c|c|c|cccccc}
		\hline\hline
		Model & MSG    & Reliabiliy      & B2 & B3 & B4 & B5 & All \\\hline
		A \cite{Jana2019} &  $C2/c'$ & GOF  & 13.17  & 11.96  & 10.36  & 9.84  & 11.02  \\
		&   &  R$_{\mathrm{p}}$~(\%) & 8.15  & 6.49  & 5.13  & 4.60  & 5.92      \\
		& &    wR$_{\mathrm{p}}$~(\%) & 9.73  & 7.57  & 5.89  & 5.29  & 7.08   \\[0.2cm]\hline
		B \cite{Ding2020} & $C2/c'$  & GOF  & 15.17  & 13.30  & 10.99  & 10.51  & 12.18  \\
		&   &  R$_{\mathrm{p}}$~(\%) & 9.26  & 7.22  & 5.47  & 5.06  & 6.57      \\
		&   &    wR$_{\mathrm{p}}$~(\%) & 11.21  & 8.42  & 6.25  & 5.65  & 7.83   \\[0.2cm]\hline
		C    & $C2/c'$  &  GOF & 7.03   & 6.85  & 7.34  & 7.15  & 7.11      \\
		& &  R$_{\mathrm{p}}$~(\%) & 4.16  & 3.66  & 3.45  & 3.15  & 3.58   \\
		& &      wR$_{\mathrm{p}}$~(\%) & 5.19  & 4.33  & 4.16  & 3.84  & 4.56   \\[0.2cm]\hline
		D    & $Cc'$     & GOF & 6.90  & 6.61  & 7.12  & 7.01  & 6.92       \\
		&     & R$_{\mathrm{p}}$~(\%) & 4.03  & 3.58  & 3.39  & 3.15  & 3.52   \\
		& &  wR$_{\mathrm{p}}$~(\%) & 5.06 & 4.15 & 4.01 & 3.73 & 4.44                 \\[0.2cm]\hline
		E    & $C2$ & GOF  & 7.00  & 6.80  & 7.34  & 7.12  & 7.05       \\
		& & R$_{\mathrm{p}}$~(\%) & 4.12  & 3.66  & 3.44  & 3.17  & 3.58   \\
		& & wR$_{\mathrm{p}}$~(\%) & 5.13 & 4.27 & 4.13 & 3.79 & 4.52         \\[0.2cm]\hline
		F & $P\bar{1}'$  & GOF & 6.88  & 6.53  & 7.09  & 6.82  & 6.86       \\
		& & R$_{\mathrm{p}}$~(\%) & 4.06  & 3.56  & 3.32  & 3.01  & 3.48   \\
		& & wR$_{\mathrm{p}}$~(\%) & 5.04 & 4.10 & 3.99 & 3.63  & 4.40               \\[0.2cm]\hline\hline
	\end{tabular}
\end{table}

\begin{figure}[t]
\includegraphics[width=\linewidth]{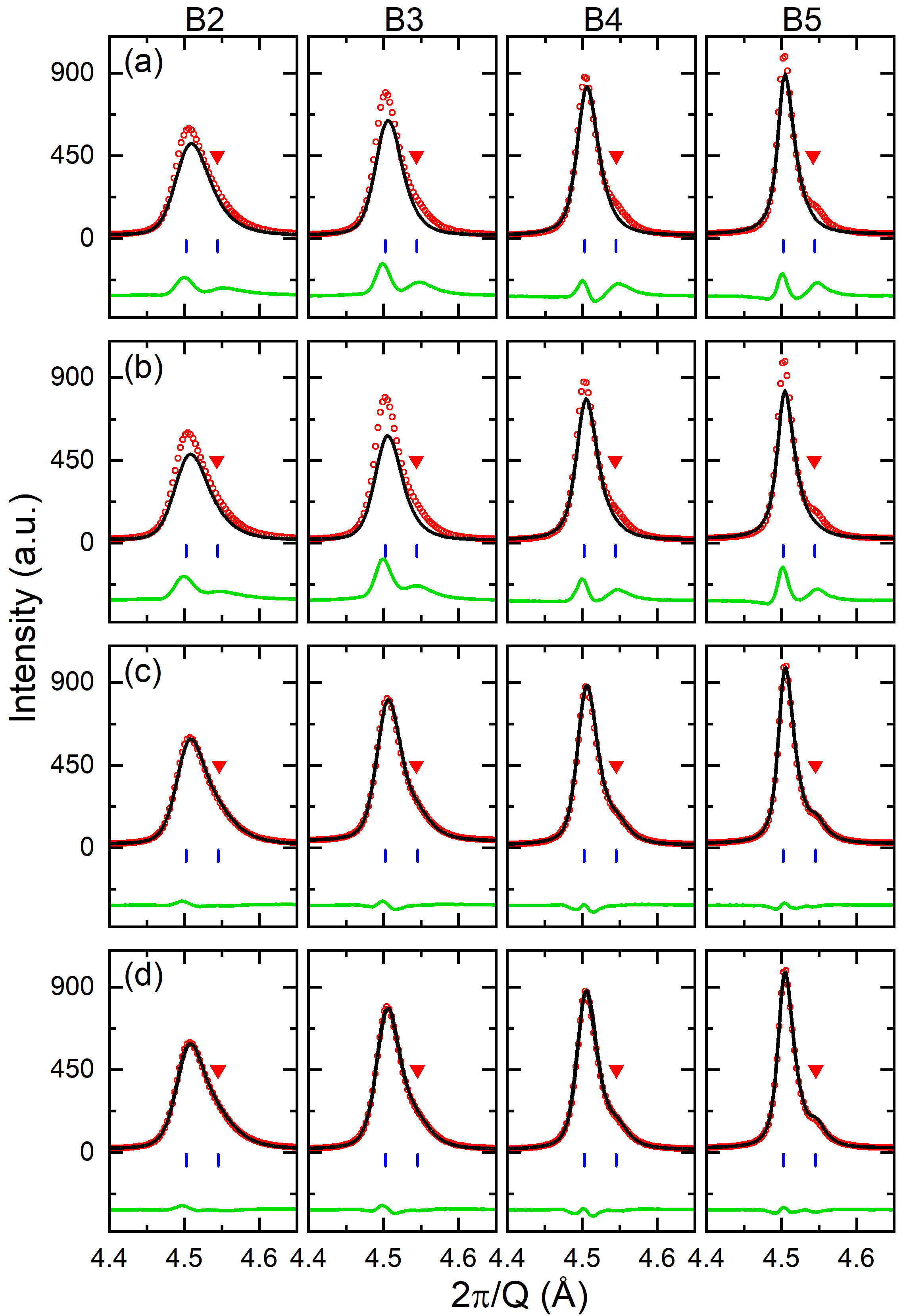}
\caption{Comparison of Rietveld refinements for the two selected peaks, (200)$_{\rm m}$ (right peak) and (110)$_{\rm m}$ (left peak). (a) Model A. (b) Model B. (c) Model C. (d) Model F. B denotes bank. The (200)$_{\rm m}$ magnetic Bragg peak (red triangles) was observed at 1.5~K on the right shoulder of the (110)$_{\rm m}$ peak, where a subscript ${\rm m}$ means monoclinic. Note that the (110) peak in $P\bar{1}'$ in (d) is based on the triclinic unit cell.
}
\label{fig:200:peak}
\end{figure}

\subsection{Successive magnetic and structural transitions upon cooling}
\label{sec:ND:TD}
Figure~\ref{fig:Block 3} presents the neutron diffraction patterns obtained from 120 to 1.5~K, focusing on the main magnetic and nuclear Bragg peaks. The top and bottom rows of Fig.~\ref{fig:Block 3} display the contour plot and stacked data, respectively. They reveal two phase transitions at 97.5~K ($T_{\rm N}$) and 87.5~K ($T_{\rm S}$), consistent with observed anomalies in the magnetic and electric measurements (Fig.~\ref{fig:Char}). At 97.5 K, the emergence of additional signals for specific Bragg peaks indicates the onset of long-range antiferromagnetic order at (004), (100), and (002) with values of 7.09, 4.52, and 3.55 \AA, respectively [Figs.~\ref{fig:Block 3}(c), \ref{fig:Block 3}(g) and \ref{fig:Block 3}(i), with the corresponding temperature-dependent data given in Figs.~\ref{fig:Block 3}(d), \ref{fig:Block 3}(h) and \ref{fig:Block 3}(j)]. In contrast, the intensities of other peaks, such as (104), (1--14), (102), and (1--12), exhibit no significant changes [Figs.~\ref{fig:Block 3}(a), ~\ref{fig:Block 3}(b), ~\ref{fig:Block 3}(e), and ~\ref{fig:Block 3}(f)].

Below 87.5~K, several Bragg peaks strongly split, except for (00\textit{l}) peaks. The previously degenerate Bragg peaks, (104) (1--14), (102), and (1-12), began to split as shown in Figs.~\ref{fig:Block 3}(a), ~\ref{fig:Block 3}(b), ~\ref{fig:Block 3}(e), and ~\ref{fig:Block 3}(f). This indicates a structural distortion in the $ab$ plane, consistent with the previously reported monoclinic structure~\cite{Jana2019, Ding2020}. Upon further cooling, magnetic Bragg peak intensities gradually increase, while nuclear peaks exhibit progressively stronger splitting. Between 87.5 and 70~K, the trigonal and monoclinic structures coexist (see the green dashed lines in the top row and the shaded areas in the bottom row of Fig.~\ref{fig:Block 3}), with decreasing trigonal peak intensity and increasing monoclinic peak intensity.

Accordingly, magnetic hysteresis is expected in the magnetization data within this temperature range due to the first-order nature of the structural transition. This observation is consistent with the emergence of the magnetic bifurcation between the ZFC and FC data below around $T_{\rm S}$ in the magnetic susceptibility [the blue lines in Fig.~\ref{fig:Char}(b) and its inset]. At 70~K, peak splitting reaches its maximum, and trigonal peaks nearly vanish, indicating the formation of a single monoclinic throughout the sample volume---a behavior that persists down to the base temperature (1.5~K). We labeled 70~K as $T_{\rm 0}$ to denote this characteristic temperature [see green lines in Fig.~\ref{fig:Block 3}]. The phase coexistence between 87.5 and 70~K corresponds to the temperature where the bigger drop ends in dielectric constants [see the dashed vertical green line in Fig.~\ref{fig:Char}(d)].

\begin{figure*}[t]
\includegraphics[width=\linewidth]{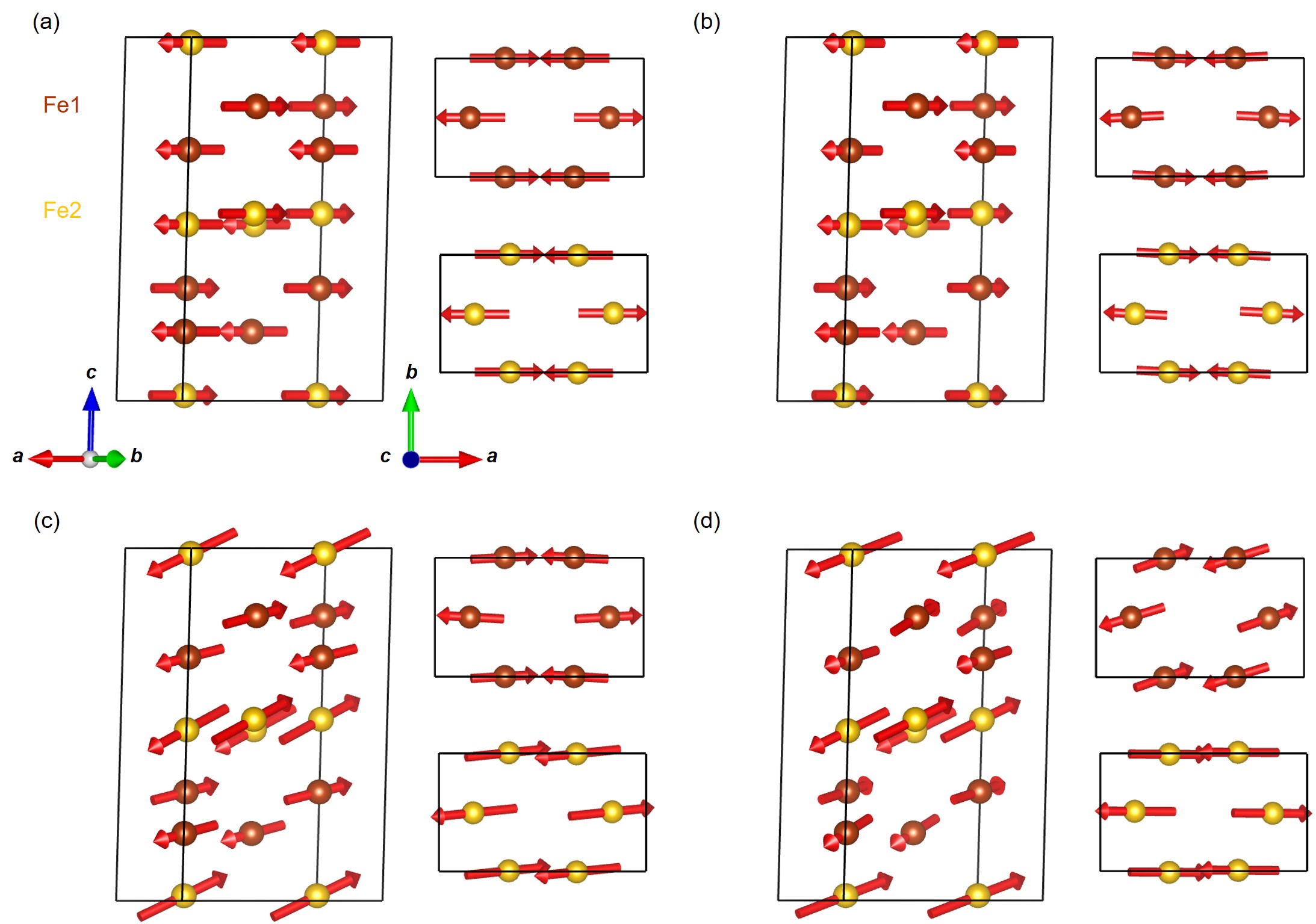}
\caption{Comparison of magnetic structural candidates refined at 1.5~K. (a) Model A. (b) Model B. (c) Model C. (d) Model F. Nb and O ions are omitted for brevity.
}
\label{fig:comp:mag:str}
\end{figure*}

\subsection{Magnetic and structural refinements}
\label{sec:ND:mag:str}
Magnetic and structural refinements were then conducted for quantitative analysis using neutron diffraction data. The nuclear structure at 120~K (well above $T_{\rm N}$) was first determined to serve as a reference for magnetic refinements at lower temperatures, consistent with the known crystal structure~\cite{Jana2019, Ding2020}. Because additional peaks, which were observed below $T_{\rm N}$, are allowed in the parent space group, we interpreted them as magnetic Bragg peaks indexed by the ordering wave vector $k$~=~(0, 0, 0). Group theory analysis using $k$ and the parent crystal structure provided candidates for MSGs (see the details in the Appendix and Fig.~\ref{fig:Group_subgroup}).

Our observations of the peak splitting align with two previous studies~\cite{Jana2019, Ding2020}, in which the structural transition from the trigonal (\sg) to the monoclinic space group ($C2/c$) was proposed to be below $T_{\rm S}$. This trend remained unchanged from 70 to 1.5~K. Thus, refinements of the magnetic and nuclear structures were refined at 1.5~K---below the base temperatures of 10~K~\cite{Jana2019} and 5~K~\cite{Ding2020} reported in the literature---to determine a more accurate magnetic ground state using high-resolution neutron diffraction data with better statistics.

Two previous studies~\cite{Jana2019, Ding2020} reported magnetic structures using $C2/c'$. Specifically, the magnetic moments of both Fe ions were assumed to lie in the $ab$ plane with the equal magnetic moment in a collinear~\cite{Jana2019} (labeled model A) or noncollinear ~\cite{Ding2020} (model B) arrangement. Therefore, we first employed their magnetic structures along with our neutron diffraction data by fixing the magnetic moments of these two models. The magnetic refinement results yielded a poor fit [Figs.~\ref{fig:mag:refine:1p5K}(a) and \ref{fig:mag:refine:1p5K}(b)], indicating that our neutron diffraction data were highly sensitive to differences between the magnetic models [see Table \ref{1p5K_magnetic_results_right_part}]. The poor refinements can be confirmed in Figs.~\ref{fig:200:peak}(a) and \ref{fig:200:peak}(a), where key peaks were compared with refinements in banks 2 to 5. Refined magnetic structures are illustrated in Figs.~\ref{fig:comp:mag:str}(a) and~\ref{fig:comp:mag:str}(b).

Subsequent refinements with freed magnetic moments led to improved results (model C), as presented in Fig.~\ref{fig:mag:refine:1p5K}(c) (see Table \ref{1p5K_magnetic_results_right_part} for comparing the \textit{R} factors). An analysis of the magnetic refinements with and without the magnetic moment component parallel to the $c$ axis ($M_{c}$) revealed a weaker magnetic Bragg peak, (200), clearly observed at \textit{d}~$\sim$~4.54~\AA~on the shoulder of a stronger peak, (110), in the monoclinic lattice. The (200) peak was better resolved in the bank 4 and 5 data due to their higher $d$-spacing resolutions. This peak was well fitted by introducing $M_{c}$ [also see Fig.~\ref{fig:mag:refine:1p5K}(c)]. A (h00) magnetic Bragg peak means that there is the magnetic moment component can lie in the \textit{b}$^{*}$\textit{c}$^{*}$ plane in the momentum space because neutrons detect only the magnetic moment perpendicular to the wave vector. As a result, we emphasize that our high-resolution neutron diffraction data revealed the (200) magnetic Bragg peak, determining a sizable $M_{c}$ component using $C2/c'$. A refined magnetic structure is illustrated in Fig.~\ref{fig:comp:mag:str}(c).

A lower MSG, $P\bar{1}'$ was also proposed in the literature based on ME measurements on \FNO single crystals~\cite{Zhang2021}. However, the magnetic structure remains unreported. Thus, we further lowered the magnetic symmetry from $C2/c'$ to $P\bar{1}'$ in a group theory analysis. The possible MSGs between them are $Cc'$ (No.~9.39) and $C2$ (No.~5.13; see Fig.~\ref{fig:Group_subgroup} for a group-subgroup diagram). Consequently, we performed magnetic refinements at 1.5~K using all three candidates, $Cc'$ (model D), $C2$ (model E), and $P\bar{1}'$ (model F). For $Cc'$, the magnetic structure obtained from $C2/c'$ was initially used with a fixed monoclinic nuclear structure~\cite{Ding2020} and the obtained magnetic moments for $C2/c'$. This exactly reproduced the same R factor as the $C2/c'$ analysis. This procedure ensured systematic and accurate structure factor calculations, even with the new MSG. The magnetic moments were then refined, followed by structural refinement. The same procedure was subsequently applied for $C2$ and $P\bar{1}'$.

We compared magnetic refinements results at 1.5~K for $C2/c'$, $Cc'$, $C2$, and $P\bar{1}'$. As compared in Table \ref{1p5K_magnetic_results_right_part}, the obtained \textit{R}$_{\mathrm{p}}$ (\textit{w}\textit{R}$_{\mathrm{p}}$) ranged from 3.48 to 3.58 (4.40 to 4.56), with a maximum difference of approximately 0.1 (0.16). As a result, the magnetic structure refined at $P\bar{1}'$ produced the best results for \textit{R}$_{\mathrm{p}}$ and \textit{w}\textit{R}$_{\mathrm{p}}$. Its magnetic structure is illustrated in Fig.~\ref{fig:comp:mag:str}(d). However, this difference could fall within the fitting uncertainty, although some studies may find them physically meaningful~\cite{Willwater2021}. Thus, we emphasize that our refinement results should be regarded as suggestive. Table~\ref{table:1p5K:magnetic:results} gives the parameters extracted from magnetic refinements. Definitions of the reliability parameters, such as \textit{R}$_{\mathrm{p}}$ and \textit{w}\textit{R}$_{\mathrm{p}}$, can be found in the literature~\cite{Yadav2023}.

The refined magnetic structures using $C2/c'$, $Cc'$, $C2$, and $P\bar{1}'$ were similar. Thus, refinement results at $C2/c'$ and $P\bar{1}'$ are shown Figs.~\ref{fig:mag:refine:1p5K}(c) and \ref{fig:mag:refine:1p5K}(d) (models C and F) as examples. The magnetic structure and the sizable magnetic moment component along the $c$ axis are common. As a result, we determined a new magnetic structure at 1.5~K using $C2/c'$ by identifying the moment along the \textit{c} axis. This magnetic structure is similar to that at $P\bar{1}'$, as compared in Figs.~\ref{fig:comp:mag:str}(c) and \ref{fig:comp:mag:str}(d). We also performed magnetic refinements without any impurity phase at 1.5~K and obtained similar results. Magnetic refinements were also repeated with the 70~K data, giving rise to similar results. Thus, we conclude that no structural or magnetic transition occurs between 70 and 1.5~K.

As a result, we present the refinement results for four different magnetic models in Figs.~\ref{fig:mag:refine:1p5K}. They compare the refinement outcomes at 1.5~K using a selective set of magnetic models. These results demonstrate that our high-resolution neutron diffraction data can distinguish and rule out two previously reported magnetic models without the $M_{c}$ component [Figs.~\ref{fig:mag:refine:1p5K}(a) and \ref{fig:mag:refine:1p5K}(b)], providing two better solutions with the $M_{c}$ moment [Figs.~\ref{fig:mag:refine:1p5K}(c) and \ref{fig:mag:refine:1p5K}(d)]. We note that the different models exhibit similar magnetic Bragg peaks. However, the predicted intensities of the Bragg peaks vary significantly, depending on the specific models, the spin moments, and the number of distinct magnetic sublattices. Figure~\ref{fig:comp:mag:str} allows us to visually inspect these differences.

Figure~\ref{fig:comp:mag:str} compares the reported magnetic structures with ours. A previous powder neutron diffraction study~\cite{Jana2019} found a collinear antiferromagnetic order at 10~K, as visualized in Fig.~\ref{fig:comp:mag:str}(a). The spins along the $c$ axis are aligned parallel and antiparallel within the $ab$ plane, forming a simple antiferromagnetic N\'{e}el phase in the honeycomb lattice stacked along the $c$ axis. The other previous study using single-crystal neutron diffraction data reported the same magnetic structure with the same $C2/c'$ MSG, except for the presence of a slight canting angle of 5.81$^{\circ}$ between the pairs of ferromagnetic chains~\cite{Ding2020} [Fig.~\ref{fig:comp:mag:str}(b)]. However, we discovered a substantially distinct magnetic order even in the same $C2'/c$ MSG [Fig.~\ref{fig:mag:refine:1p5K}(c)]. The magnetic moments in the two sublattices were unequal, which may be more realistic given their chemically distinct positions, different Wyckoff sites, and varying oxygen environments. The magnetic moments of the flat layer (Fe1) were consistently smaller than those of the buckled layer (Fe2) across models. The refined magnetic structure with $P\bar{1}'$ is shown in Fig.~\ref{fig:comp:mag:str}(d) for comparison.

\subsection{{\it Ab initio} calculations}
\label{sec:abinitio}
We further performed {\it ab initio} calculations to compare the energy of the six magnetic models (A to F; see Table~\ref{table:1p5K:magnetic:results}). All crystal structures were fixed to that of model F to provide a consistent basis for comparing the energies of the models. The unit cell size differs only slightly among the six models. The magnetic configurations were introduced based on the experimental results (Table~\ref{table:1p5K:magnetic:results}) by initializing the Fe magnetic moments. During the self-consistent field iterations, the magnetic moment directions initialized for each configuration remained stable, indicating that the energy landscape respects the imposed magnetic symmetries.

\begin{figure}[t]
\includegraphics[width=\linewidth]{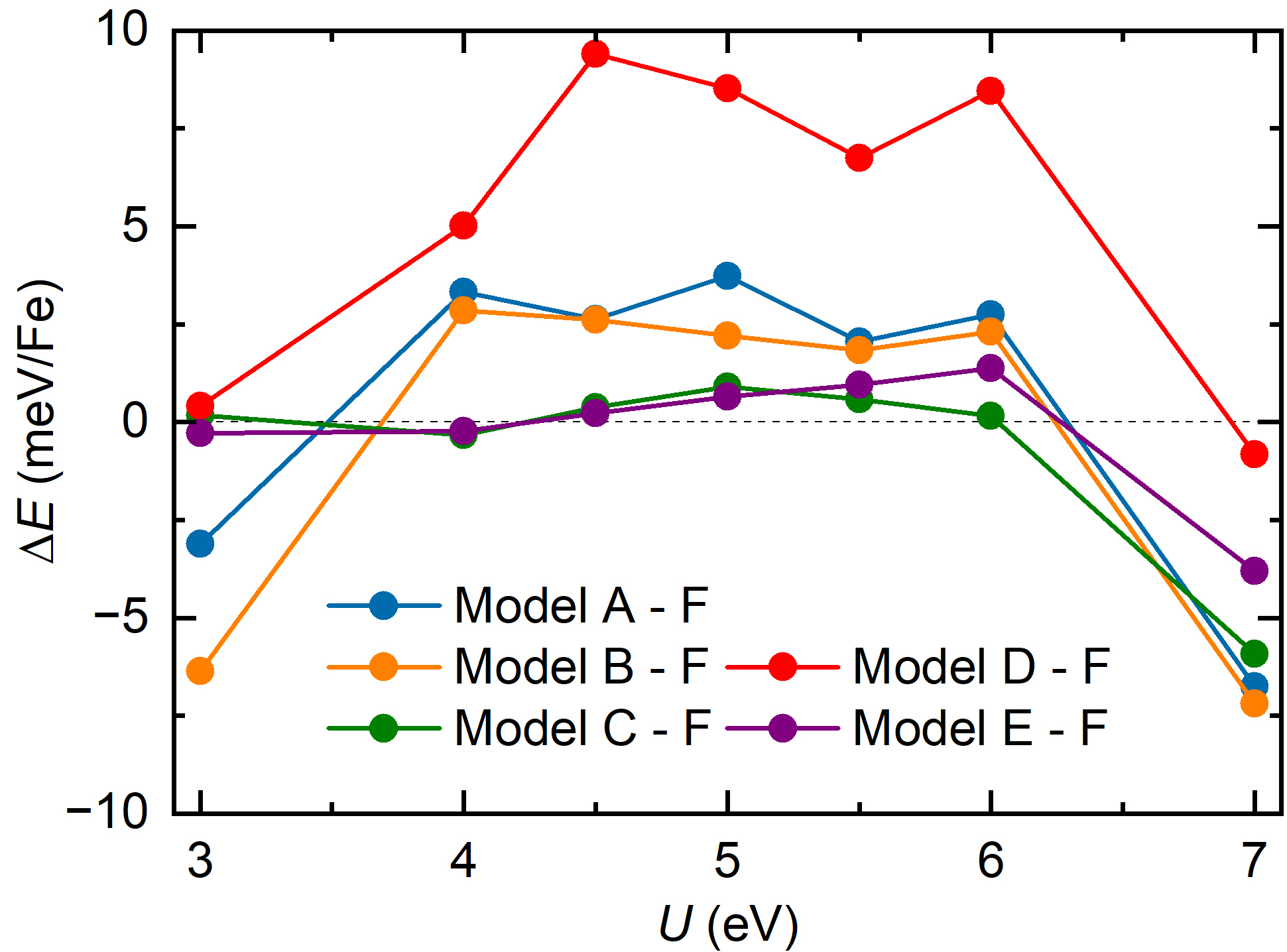}
\caption{Comparison of the differences in the calculated total energy between magnetic models.
}
\label{fig:comp:dft}
\end{figure}

Figure~\ref{fig:comp:dft} shows the total energy differences using the energy of model F subtracted as the reference. Model F is found to be the energetically most favorable configuration within the range of $U_{\rm eff}$ = 4.5 to 6.0 eV, which is a sensible range for the Fe$^{2+}$ correlation energy, with similar energies in models C, E, and F. This theoretical result is consistent with a small difference in \textit{R} factor values from experiments, as shown in Fig.~\ref{fig:mag:refine:1p5K} and Table~\ref{1p5K_magnetic_results_right_part}. The relatively lower total energies of models C to F in comparison to those of models A and B suggest that magnetic configurations featuring the finite out-of-plane spin component are energetically preferred in the calculation, too.

\section{Discussion}
\label{sec:discussion}
In this work, we determined a new magnetic structure of \FNO at 1.5~K using the $C2/c'$ MSG and then extended the analysis down to the highest possible ($P\bar{1}'$) MSG compatible with all nine ME components. Our results motivate a discussion of the potential microscopic mechanisms underlying the ME effect in \FNO. The literature identifies three common mechanisms to explain magnetically induced ferroelectricity~\cite{Tokura2014}: spin current~\cite{Katsura2005}, exchange striction~\cite{Sergienko2006}, and \textit{p}-\textit{d} hybridization~\cite{Arima2007}.

A previous study observed an anisotropic ME effect and electric polarization reversal, explained by a metal-ligand \textit{p}-\textit{d} hybridization mechanism in the \sg~and $C2/c$ space group~\cite{Zhang2023}. However, this mechanism fails to explain the net electric polarization under $H \parallel c$, as the in-plane magnetic moments were still antiparallel in the two previous magnetic structural candidates~\cite{Jana2019, Ding2020}. This issue can be resolved using our magnetic models where the in-plane spins do not cancel under $H \parallel c$ (models C and F). Also, the ME tensor of the $C2/c'$ MSG at both ambient and high-field states~\cite{Choi2020} cannot explain the observed electric polarization for all nine components~\cite{Zhang2021}. On the other hand, for the magnetic structure using $P\bar{1}'$, all nine components of the ME tensor are possible by symmetry even at ambient magnetic fields. This might be consistent with the electric polarization experimentally observed under the strong magnetic field for all nine directions~\cite{Zhang2023}. The ME tensors of the $C2/c'$ and $P\bar{1}'$ MSGs are compared in Table~\ref{tab:ME}. If $C2/c'$ is chosen as the magnetic ground state of \FNO, the additional ME tensor components of $\alpha_{12}$, $\alpha_{21}$, $\alpha_{23}$, and $\alpha_{32}$ cannot be explained. However, revisited ME measurements on single crystals without or with weak magnetic fields will be needed to confirm the ambient magnetic symmetry. The ME analysis based on the magnetic structure established in this study could advance the understanding of its unconventional ME properties in future studies.

\begin{table}
	\caption{The ME tensors ($\alpha_{ij}$) of the $C2/c'$ and $P\bar{1}'$. The ME tensor is defined as $P_{i}$ = $\alpha_{\textit{ij}}$ $H_{j}$ where $P_{i}$ ($H_{j}$) is the electric polarization (the magnetic field) along the three principal directions $i$ ($j$).}
	\label{tab:ME}
	\begin{center}
		\setlength{\tabcolsep}{16pt}
		\renewcommand{\arraystretch}{1.5}
		\begin{tabular}{|c|c|}
			\hline
			$C2/c'$ (No.~15.88) & $P\bar{1}'$ (No.~2.6) \\
			\hline
			\rule{0pt}{3ex}
			$\left[
			\begin{array}{ccc}
				\alpha_{11} & 0 & \alpha_{13} \\
				0 & \alpha_{22} & 0 \\
				\alpha_{31} & 0 & \alpha_{33}
			\end{array}
			\right]$
			\rule[-1.5ex]{0pt}{0pt}
			&
			\rule{0pt}{7ex}
			$\left[
			\begin{array}{ccc}
				\alpha_{11} & \alpha_{12} & \alpha_{13} \\
				\alpha_{21} & \alpha_{22} & \alpha_{23} \\
				\alpha_{31} & \alpha_{32} & \alpha_{33}
			\end{array}
			\right]$
			\rule[-1.5ex]{0pt}{0pt} \\
			\hline
		\end{tabular}
	\end{center}
\end{table}

The structure using $P\bar{1}'$ has been presented through a comprehensive experimental and theoretical analysis in this study. However, we emphasize that the qualitative differences in the reliability factors compared to those of $C2/c'$ are minimal in terms of refinement results. Additionally, the difference in the calculated ground state energies between these two magnetic space groups is negligible (Fig.~\ref{fig:comp:dft}). To further verify the magnetic ground state and distinguish between these two MSGs, it will be crucial to observe the peak splitting in the triclinic structure experimentally. In this regard, high-resolution x-ray diffraction could be useful in future studies.

Compared to \FNO, the related compound \FTO has received less attention despite the proposed coexistence of linear magnetoelectricity and type-II multiferroism~\cite{Maignan2018:2}. Similarly, based on our results for \FNO, \FTO may exhibit low magnetic symmetry capable of explaining its unusual ME properties. Our results also suggests the possibility of similar behavior in the related compound. This study underscores the need for careful investigation of the \ABO compound family to establish more accurate magnetic structures, which are essential for understanding their unconventional ME properties.

\section{Conclusions}
\label{sec:conclusions}
This study established a more realistic magnetic structure for honeycomb magnetoelectric \FNO through powder neutron diffraction. High-quality data combined with group analysis enabled us to determine the out-of-plane magnetic moment perpendicular to the honeycomb layer. Previously reported magnetic structures within the same $C2/c'$ MSG without the $ M_c$ component were not fitted with our data. Thus, they can be ruled out as candidates for the magnetic ground state. Furthermore, motivated by the experimental observation of the complete ME tensor components, we provided the magnetic structure in a lower MSG, $P\bar{1}'$, where the $M_{c}$ is also substantial. This study provides key information---the MSG and its symmetry---crucial for understanding unusually strong ME properties of honeycomb noncollinear antiferromagnetic \FNO.

\begin{table*} [t]
	\caption{Magnetic and nuclear reﬁnement results at 1.5~K for magnetic space groups $C2/c'$, $Cc'$, $C2$, and $P\bar{1}'$. The unit of the fitted moments is $\mu_\textrm{B}$.
	}
	\vspace{0.2cm}
	\label{table:1p5K:magnetic:results}
	\setlength\extrarowheight{4pt}
	\setlength{\tabcolsep}{4pt}
	\begin{tabular}{c|c|c|ccccccc}
		\hline\hline
		& MSG & Atoms & x & y & z  & $M_{x}$    & $M_{y}$    & $M_{z}$     & M   \\\hline
		A \cite{Jana2019} &  $C2/c'$  & Fe1 & 0.83415(16) &   0.4989(5) &   0.69045(9)  & 3.83  & 0  & 0  & 3.83     \\
		&   & Fe2 & 0.83565(14) &   0.5046(5)  &  0.48511(7)   & 3.83  & 0  & 0  & 3.83             \\[0.2cm]\hline
		
		B \cite{Ding2020} & $C2/c'$  & Fe1 & 0.83021(18)  &  0.4981(4)  &  0.68867(10)  &  3.52  & 0.18  & 0  & 3.52      \\
		&   & Fe2 & 0.83181(16)  &  0.5038(5) &   0.48383(7)    & 3.52  & -0.18  & 0  & 3.52         \\[0.2cm]\hline
		
		C    & $C2/c'$  & Fe1 & 0.83410(14) & 0.4985(4) & 0.69111(9)  & 3.651(4)  & -0.204(8)  & -0.863(11)  & 3.782(14)      \\
		& & Fe2 & 0.83585(12) & 0.5044(3) & 0.48523(6)  & 4.675(11)  & 0.484(9)   & -2.151(15)   & 5.226(21)      \\[0.2cm]\hline
		
		D     & $Cc'$  & Fe1\_1  & 0.83247(15) & 0.5018(11) & 0.6926(2)  & 3.913(38)  & -0.403(104)  & 0.027(47)  & 3.933(120)       \\
		& & Fe1\_2 & 0.16757(15) & 0.4993(13) & 0.8094(2) & -2.998(36)  & 0.011(60)   & 2.338(43)   & 3.856(83)    \\
		& & Fe2\_1 & 0.83434(14) & 0.4998(10) & 0.48486(18)   & 5.839(41)  & 0.558(65)   & -2.155(46)   & 6.309(90)     \\	
		& & Fe2\_2 & 0.16519(15) & 0.5036(12) & 0.0147(2)   & -3.577(31)  & 0.264(40)   & 1.888(44)   & 4.102(67)   \\[0.2cm]\hline
		
		E     & $C2$ & Fe1\_1 & 0.8368(4) & 0.49949(6) & 0.6907(3)  & 3.672(52)  & -0.293(68)  & -1.101(60)  & 3.876(104)      \\
		& & Fe1\_2 & 0.1723(5) & 0.50053(6) & 0.3088(3)  & -3.186(51)  & -0.683(71)   & 1.219(62)   & 3.512(107)       \\
		& & Fe2\_1 & 0.8370(4) & 0.50013(6) & 0.4839(3)   & 4.681(47)  & 0.420(72)   & -2.334(43)   & 5.308(96)          \\	
		& & Fe2\_2 & 0.1657(4) & 0.49982(6) & 0.5150(3)   & -5.057(48)  & 0.978(70)   & 1.578(40)   & 5.431(94)   \\[0.2cm]\hline
		
		F & $P\bar{1}'$   & Fe1\_1  & 0.8320(4) & 0.4948(7) & 0.6887(2)  & 3.575(24)  & 1.098(99)  & -0.865(44)  & 3.863(111)        \\
		& & Fe1\_2 & 0.1706(3) & 0.5028(7) & 0.8070(2)  & -3.209(25)  & -1.122(97)   & 1.481(39)   & 3.746(107)   \\
		& & Fe2\_1 & 0.8337(3) & 0.4915(7) & 0.4840(2)  & 4.369(35)  & -0.033(121)   & -1.939(38)   & 4.832(132)   \\	
		& & Fe2\_2 & 0.1647(3) & 0.5094(6) & 0.0143(2)   & -5.431(35)  & -0.049(120)   & 1.874(35)   & 5.798(130) \\[0.2cm]\hline\hline
	\end{tabular}
\end{table*}

\begin{table*}
	\caption{Symmetry relations of the two magnetic space group candidates for \FNO. The Fe ions are given in Table~\ref{table:1p5K:magnetic:results}.
	}
	\vspace{0.1cm}
	\label{tab_sym}
	\begin{center}
		\resizebox{\textwidth}{!}{
			\setlength\extrarowheight{4pt}
			\begin{tabular}
				[c]{c|c|c|c|c|c}
				\hline\hline
				Model &	MSG	& Atoms  &    Wyckoff position &       Multiplicity            & Magnetic moment \\
				\hline
				C	& $C2/c'$	& Fe1  & ($x$, $y$, $z$ $|$ $m_{x}$, $m_{y}$, $m_{z}$) (-$x$, $y$, -$z$+1/2 $|$ -$m_{x}$, $m_{y}$, -$m_{z}$)       &  8         &  ($M_{x}$, $M_{y}$, $M_{z}$)                                  \\
				&	&     & (-$x$, -$y$, -$z$ $|$ -$m_{x}$, -$m_{y}$, -$m_{z}$) ($x$, -$y$, $z$+1/2 $|$ $m_{x}$, -$m_{y}$, $m_{z}$)        &          &   \\
				&	 & & ($x$+1/2, $y$+1/2, $z$ $|$ $m_{x}$, $m_{y}$, $m_{z}$) (-$x$+1/2, $y$+1/2, -$z$+1/2 $|$ -$m_{x}$, $m_{y}$, -$m_{z}$) &   & \\
				&	 & & (-$x$+1/2, -$y$+1/2, -$z$ $|$ -$m_{x}$, -$m_{y}$, -$m_{z}$) ($x$+1/2, -$y$+1/2, $z$+1/2 $|$ $m_{x}$, -$m_{y}$, $m_{z}$) &  & \\
				\cline{3-6}	
				&	& Fe2  & ($x$, $y$, $z$ $|$ m$_{x}$, $m_{y}$, $m_{z}$) (-$x$, $y$, -$z$+1/2 $|$ -$m_{x}$, $m_{y}$, -$m_{z}$)       &   8        &    ($M_{x}$, $M_{y}$, $M_{z}$)                                \\
				&	  &   & (-$x$, -$y$, -$z$ $|$ -$m_{x}$, -$m_{y}$, -$m_{z}$) ($x$, -$y$, $z$+1/2 $|$ $m_{x}$, -$m_{y}$, $m_{z}$)        &          &   \\
				&	 & & ($x$+1/2, $y$+1/2, $z$ $|$ $m_{x}$, $m_{y}$, $m_{z}$) (-$x$+1/2, $y$+1/2, -$z$+1/2 $|$ -$m_{x}$, $m_{y}$, -$m_{z}$) &   & \\
				&	 & & (-$x$+1/2, -$y$+1/2, -$z$ $|$ -$m_{x}$, -$m_{y}$, -$m_{z}$) ($x$+1/2, -$y$+1/2, $z$+1/2 $|$ $m_{x}$, -$m_{y}$, $m_{z}$) &  & \\
				
				\hline
				F	&	$P\bar{1}'$	& Fe1\_1  & ($x$, $y$, $z$ $|$ $m_{x}$, $m_{y}$, $m_{z}$) (-$x$, -$y$, -$z$ $|$ -$m_{x}$, -$m_{x}$, -$m_{z}$)   &   4           &   ($M_{x}$, $M_{y}$, $M_{z}$)                                     \\
				&		 &  & ($x$+1/2, $y$+1/2, $z$ $|$ $m_{x}$, $m_{y}$, $m_{z}$) (-$x$+1/2, -$y$+1/2, -$z$ $|$ -$m_{x}$, -$m_{y}$, -$m_{z}$)        &     &              \\
				\cline{3-6}
				&		& Fe1\_2  & (-$x$, $y$, -$z$+1/2 $|$ $m_{x}$, $m_{y}$, $m_{z}$) ($x$, -$y$, $z$+1/2 $|$ -$m_{x}$, -$m_{y}$, -$m_{z}$) & 4 & ($M_{x}$, $M_{y}$, $M_{z}$)  \\
				&	  &   & (-$x$+1/2, $y$+1/2, -$z$+1/2 $|$ $m_{x}$, $m_{y}$, $m_{z}$) ($x$+1/2, -$y$+1/2, $z$+1/2 $|$ -$m_{x}$, -$m_{y}$, -$m_{z}$) &  &   \\
				\cline{3-6}
				&	& Fe2\_1	   & ($x$, $y$, $z$ $|$ $m_{x}$, $m_{y}$, $m_{z}$) (-$x$, -$y$, -$z$ $|$ -$m_{x}$, -$m_{x}$, -$m_{z}$)   &  4            &  ($M_{x}$, $M_{y}$, M$_{z}$)                                      \\
				&	  &  & ($x$+1/2, $y$+1/2, $z$ $|$ $m_{x}$, $m_{y}$, $m_{z}$) (-$x$+1/2, -$y$+1/2, -$z$ $|$ -$m_{x}$, -$m_{y}$, -$m_{z}$)        &     &              \\
				\cline{3-6}
				&	& Fe2\_2	 & (-$x$, $y$, -$z$+1/2 $|$ $m_{x}$, $m_{y}$, $m_{z}$) ($x$, -$y$, $z$+1/2 $|$ -$m_{x}$, -$m_{y}$, -$m_{z}$) & 4 & ($M_{x}$, $M_{y}$, $M_{z}$)  \\
				&	  &  & (-$x$+1/2, $y$+1/2, -$z$+1/2 $|$ $m_{x}$, $m_{y}$, $m_{z}$) ($x$+1/2, -$y$+1/2, $z$+1/2 $|$ -$m_{x}$, -$m_{y}$, -$m_{z}$) &  &    \\

				\hline\hline
			\end{tabular}
		}
	\end{center}
\end{table*}

\begin{table}[]
	\caption{List of Bragg peaks at 70~K for $P\bar{1}'$ presented in Fig.~\ref{fig:Block 3}.
	}
	\vspace{0.2cm}
	\label{Table:C6}
	\setlength\extrarowheight{4pt}
	\setlength{\tabcolsep}{10pt}
	\begin{tabular}{ccc|c|c}
		\hline
		\hline
		h & k  & l  & \textit{d} spacing (\AA) & Figure\\\hline
		
		0 & 0  & 2  & 7.10749   &   \ref{fig:Block 3}(j)           \\
		\hline
		2 & 0  & 0  & 4.53899	&	\ref{fig:Block 3}(h)			\\
		1 & 1  & 0  & 4.50927	&				\\
		1 & -1  & 0  & 4.50809	&				\\
		\hline
		-2 &  0  & 2  & 3.86656 	&	\ref{fig:Block 3}(f)			\\
		-1 &  -1  & 2  & 3.82852 &					\\
		-1 &  1  & 2  & 3.82382	&				\\
		1 & -1  & 2  & 3.7856	&				\\
		1 &  1  & 2  & 3.78245	&			\\
		2 &  0  & 2  & 3.78104	&				\\
		\hline
		0 &  0  & 4  & 3.54825	&	\ref{fig:Block 3}(d)			\\
		\hline	
		0 &  0  & 5  & 2.83801 	&	\ref{fig:Block 3}(b)			\\
		-2 &  0  & 4  & 2.82844	&				\\
		-1 & -1  & 4  & 2.80575 	&				\\
		-1 &  1  & 4  & 2.80234	&				\\
		1 & -1  & 4  & 2.77229	&				\\
		1 & 1  & 4  & 2.76954	&				\\
		2 &  0  & 4  & 2.76165	&				\\
		\hline
		\hline
	\end{tabular}
\end{table}

\section{Acknowledgments}
We thank P. Manuel for helping with the neutron experiment and careful comments on the paper. This work was supported by the Institute for Basic Science (Grants No. IBS-R011-Y3 and No. IBS-R034-Y1) and the Advanced Facility Center for Quantum Technology at Sungkyunkwan University. Part of this study was performed using facilities at the IBS Center for Correlated Electron Systems, Seoul National University. Work at Rutgers (V.K.) was supported by the DOE under Grant No. DE-FG02-07ER46382. The work at Yonsei was supported by the National Research Foundation of Korea (NRF) through Grants No. RS-2021-NR058241 and No. RS-2022-NR069338. D.K. and A.G. are supported by Global-Learning \& Academic research institution for Master's $\cdot$ Ph.D. students, and Postdocs (G-LAMP) Program of the National Research Foundation of Korea (NRF) grant funded by the Ministry of Education (Grant No. RS-2024-00442775). A.G. is supported by the NRF (Grants No. 2021R1C1C1010429, No. 2023M3K5A1094813, and No. RS-2025-00515360). The neutron diffraction (WISH, Proposal No. RB1910501~\cite{Choi2019}) presented in this work is available in the ISIS database.

\section{Data availability}
The data that support the findings of this article are openly available~\cite{Choi2019}.

\appendix

\setcounter{figure}{0}
\setcounter{table}{0}
\setcounter{section}{0}
\renewcommand\thefigure{A\arabic{figure}}
\renewcommand\thetable{A\arabic{table}}
\renewcommand{\thesection}{A\arabic{section}}

\section{Group theory analysis}
\label{app:MSG}
In this Appendix, we provide the details of the group theory analysis. A group-subgroup symmetry analysis was conducted utilizing the K-SUBGROUPSMAG program from the Bilbao Crystallographic Server~\cite{Aroyo2011}, in conjunction with JANA2006~\cite{Petricek2014}. Due to a structural phase transition that occurs upon cooling, characterized by peak splitting, the parent trigonal structure transitions to a monoclinic lattice. Consequently, the analysis was performed using the low-temperature monoclinic structure, which was previously reported at 40 K~\cite{Ding2020}. A group-subgroup diagram was obtained using the crystal structure and the propagation wave vector $k$ = (0, 0, 0) (Fig.~\ref{fig:Group_subgroup}). Among the $C2/c'$ (No.~15.88) and $C2'/c$ (No.~15.87) MSGs, we found that the former presented a good fit with the data collected at 1.5 K.

To further investigate the potential for lower symmetries, we also conducted magnetic and structural refinements using subgroups of $C2/c'$, such as $Cc'$, $C2$, and $P\bar{1}'$, as illustrated in Fig.~\ref{fig:mag:refine:1p5K}. The refinement results are presented in Table~\ref{table:1p5K:magnetic:results}, where the Fe ions are divided into two independent sites for the subgroups of $C2/c'$. Further, Table~\ref{tab_sym} explicitly shows the magnetic symmetry relations for two candidate MSGs, $C2/c'$ and $P\bar{1}'$, obtained from the monoclinic lattice. Given the symmetries, a finite $M_{c}$ is allowed for both subgroups. Note that the same subgroup tree is derived when utilizing the parent trigonal structure; however, the number of atoms is halved since it started from the trigonal lattice.

The $P1$ MSG, a subgroup of $P\bar{1}'$, is also feasible based on our group analysis, as shown in Fig.~\ref{fig:Group_subgroup}. It also accommodates all nine magnetoelectric components. However, in this study, we selected $P\bar{1}'$ as the magnetic ground state due to its higher symmetry. This approach is standard for determining the magnetic space group, allowing a simpler explanation of the material with a larger number of available symmetry operators. The $P\bar{1}'$ MSG is the highest symmetric candidate accommodating all nine ME components, based on our group theory analysis, as shown in Fig.~\ref{fig:Group_subgroup}. The peak positions and indexing at 70 K for $P\bar{1}'$, as shown in Fig.~\ref{fig:Block 3}, are given in Table~\ref{Table:C6} for completeness.

\bibliography{Ref}

\begin{thebibliography}{40}%
\makeatletter
\providecommand \@ifxundefined [1]{%
 \@ifx{#1\undefined}
}%
\providecommand \@ifnum [1]{%
 \ifnum #1\expandafter \@firstoftwo
 \else \expandafter \@secondoftwo
 \fi
}%
\providecommand \@ifx [1]{%
 \ifx #1\expandafter \@firstoftwo
 \else \expandafter \@secondoftwo
 \fi
}%
\providecommand \natexlab [1]{#1}%
\providecommand \enquote  [1]{``#1''}%
\providecommand \bibnamefont  [1]{#1}%
\providecommand \bibfnamefont [1]{#1}%
\providecommand \citenamefont [1]{#1}%
\providecommand \href@noop [0]{\@secondoftwo}%
\providecommand \href [0]{\begingroup \@sanitize@url \@href}%
\providecommand \@href[1]{\@@startlink{#1}\@@href}%
\providecommand \@@href[1]{\endgroup#1\@@endlink}%
\providecommand \@sanitize@url [0]{\catcode `\\12\catcode `\$12\catcode
  `\&12\catcode `\#12\catcode `\^12\catcode `\_12\catcode `\%12\relax}%
\providecommand \@@startlink[1]{}%
\providecommand \@@endlink[0]{}%
\providecommand \url  [0]{\begingroup\@sanitize@url \@url }%
\providecommand \@url [1]{\endgroup\@href {#1}{\urlprefix }}%
\providecommand \urlprefix  [0]{URL }%
\providecommand \Eprint [0]{\href }%
\providecommand \doibase [0]{https://doi.org/}%
\providecommand \selectlanguage [0]{\@gobble}%
\providecommand \bibinfo  [0]{\@secondoftwo}%
\providecommand \bibfield  [0]{\@secondoftwo}%
\providecommand \translation [1]{[#1]}%
\providecommand \BibitemOpen [0]{}%
\providecommand \bibitemStop [0]{}%
\providecommand \bibitemNoStop [0]{.\EOS\space}%
\providecommand \EOS [0]{\spacefactor3000\relax}%
\providecommand \BibitemShut  [1]{\csname bibitem#1\endcsname}%
\let\auto@bib@innerbib\@empty
\bibitem [{\citenamefont {Cheong}\ and\ \citenamefont
  {Mostovoy}(2007)}]{Cheong2007}%
  \BibitemOpen
  \bibfield  {author} {\bibinfo {author} {\bibfnamefont {S.-W.}\ \bibnamefont
  {Cheong}}\ and\ \bibinfo {author} {\bibfnamefont {M.}~\bibnamefont
  {Mostovoy}},\ }\bibfield  {title} {\bibinfo {title} {{Multiferroics: A
  magnetic twist for ferroelectricity}},\ }\href
  {https://doi.org/10.1038/nmat1804} {\bibfield  {journal} {\bibinfo  {journal}
  {Nature Materials}\ }\textbf {\bibinfo {volume} {6}},\ \bibinfo {pages} {13}
  (\bibinfo {year} {2007})}\BibitemShut {NoStop}%
\bibitem [{\citenamefont {Spaldin}\ and\ \citenamefont
  {Ramesh}(2019)}]{Spaldin2019}%
  \BibitemOpen
  \bibfield  {author} {\bibinfo {author} {\bibfnamefont {N.~A.}\ \bibnamefont
  {Spaldin}}\ and\ \bibinfo {author} {\bibfnamefont {R.}~\bibnamefont
  {Ramesh}},\ }\bibfield  {title} {\bibinfo {title} {{Advances in
  magnetoelectric multiferroics}},\ }\href
  {https://doi.org/10.1021/acs.cgd.0c00679} {\bibfield  {journal} {\bibinfo
  {journal} {Nature Materials}\ }\textbf {\bibinfo {volume} {18}},\ \bibinfo
  {pages} {203} (\bibinfo {year} {2019})}\BibitemShut {NoStop}%
\bibitem [{\citenamefont {Eerenstein}\ \emph {et~al.}(2006)\citenamefont
  {Eerenstein}, \citenamefont {Mathur},\ and\ \citenamefont
  {Scott}}]{Eerenstein2006}%
  \BibitemOpen
  \bibfield  {author} {\bibinfo {author} {\bibfnamefont {W.}~\bibnamefont
  {Eerenstein}}, \bibinfo {author} {\bibfnamefont {N.}~\bibnamefont {Mathur}},\
  and\ \bibinfo {author} {\bibfnamefont {J.~F.}\ \bibnamefont {Scott}},\
  }\bibfield  {title} {\bibinfo {title} {{Multiferroic and magnetoelectric
  materials}},\ }\href {https://doi.org/https://doi.org/10.1038/nature05023}
  {\bibfield  {journal} {\bibinfo  {journal} {Nature}\ }\textbf {\bibinfo
  {volume} {442}},\ \bibinfo {pages} {759} (\bibinfo {year}
  {2006})}\BibitemShut {NoStop}%
\bibitem [{\citenamefont {Dzyaloshinskii}(1960)}]{Dzyaloshinskii1960}%
  \BibitemOpen
  \bibfield  {author} {\bibinfo {author} {\bibfnamefont {I.~E.}\ \bibnamefont
  {Dzyaloshinskii}},\ }\bibfield  {title} {\bibinfo {title} {{On the
  Magneto-Electrical Effects in Antiferromagnets}},\ }\href
  {http://www.jetp.ras.ru/cgi-bin/index/e/10/3/p628?a=list} {\bibfield
  {journal} {\bibinfo  {journal} {Journal of Experimental and Theoretical
  Physics (U.S.S.R.)}\ }\textbf {\bibinfo {volume} {10}},\ \bibinfo {pages}
  {628} (\bibinfo {year} {1960})}\BibitemShut {NoStop}%
\bibitem [{\citenamefont {Astrov}(1960)}]{Astrov1960}%
  \BibitemOpen
  \bibfield  {author} {\bibinfo {author} {\bibfnamefont {D.}~\bibnamefont
  {Astrov}},\ }\bibfield  {title} {\bibinfo {title} {{The Magnetoelectric
  Effect in Antiferromagnetics}},\ }\href
  {http://jetp.ras.ru/cgi-bin/e/index/e/11/3/p708?a=list} {\bibfield  {journal}
  {\bibinfo  {journal} {Journal of Experimental and Theoretical Physics}\
  }\textbf {\bibinfo {volume} {11}},\ \bibinfo {pages} {708} (\bibinfo {year}
  {1960})}\BibitemShut {NoStop}%
\bibitem [{\citenamefont {Tokura}\ \emph {et~al.}(2014)\citenamefont {Tokura},
  \citenamefont {Seki},\ and\ \citenamefont {Nagaosa}}]{Tokura2014}%
  \BibitemOpen
  \bibfield  {author} {\bibinfo {author} {\bibfnamefont {Y.}~\bibnamefont
  {Tokura}}, \bibinfo {author} {\bibfnamefont {S.}~\bibnamefont {Seki}},\ and\
  \bibinfo {author} {\bibfnamefont {N.}~\bibnamefont {Nagaosa}},\ }\bibfield
  {title} {\bibinfo {title} {{Multiferroics of spin origin}},\ }\href
  {https://doi.org/10.1088/0034-4885/77/7/076501} {\bibfield  {journal}
  {\bibinfo  {journal} {Reports on Progress in Physics}\ }\textbf {\bibinfo
  {volume} {77}},\ \bibinfo {pages} {076501} (\bibinfo {year}
  {2014})}\BibitemShut {NoStop}%
\bibitem [{\citenamefont {Kimura}\ \emph {et~al.}(2003)\citenamefont {Kimura},
  \citenamefont {Goto}, \citenamefont {Shintani}, \citenamefont {Ishizaka},
  \citenamefont {Arima},\ and\ \citenamefont {Tokura}}]{Kimura2003}%
  \BibitemOpen
  \bibfield  {author} {\bibinfo {author} {\bibfnamefont {T.}~\bibnamefont
  {Kimura}}, \bibinfo {author} {\bibfnamefont {T.}~\bibnamefont {Goto}},
  \bibinfo {author} {\bibfnamefont {H.}~\bibnamefont {Shintani}}, \bibinfo
  {author} {\bibfnamefont {K.}~\bibnamefont {Ishizaka}}, \bibinfo {author}
  {\bibfnamefont {T.}~\bibnamefont {Arima}},\ and\ \bibinfo {author}
  {\bibfnamefont {Y.}~\bibnamefont {Tokura}},\ }\bibfield  {title} {\bibinfo
  {title} {{Magnetic control of ferroelectric polarization}},\ }\href
  {https://doi.org/10.1038/nature02018} {\bibfield  {journal} {\bibinfo
  {journal} {Nature}\ }\textbf {\bibinfo {volume} {426}},\ \bibinfo {pages}
  {55} (\bibinfo {year} {2003})}\BibitemShut {NoStop}%
\bibitem [{\citenamefont {Hur}\ \emph {et~al.}(2004)\citenamefont {Hur},
  \citenamefont {Park}, \citenamefont {Sharma}, \citenamefont {Ahn},
  \citenamefont {Guha},\ and\ \citenamefont {Cheong}}]{Hur2004}%
  \BibitemOpen
  \bibfield  {author} {\bibinfo {author} {\bibfnamefont {N.}~\bibnamefont
  {Hur}}, \bibinfo {author} {\bibfnamefont {S.}~\bibnamefont {Park}}, \bibinfo
  {author} {\bibfnamefont {P.~A.}\ \bibnamefont {Sharma}}, \bibinfo {author}
  {\bibfnamefont {J.}~\bibnamefont {Ahn}}, \bibinfo {author} {\bibfnamefont
  {S.}~\bibnamefont {Guha}},\ and\ \bibinfo {author} {\bibfnamefont {S.-W.}\
  \bibnamefont {Cheong}},\ }\bibfield  {title} {\bibinfo {title} {{Electric
  polarization reversal and memory in a multiferroic material induced by
  magnetic fields}},\ }\href {https://doi.org/10.1038/nature02572} {\bibfield
  {journal} {\bibinfo  {journal} {Nature}\ }\textbf {\bibinfo {volume} {429}},\
  \bibinfo {pages} {392} (\bibinfo {year} {2004})}\BibitemShut {NoStop}%
\bibitem [{\citenamefont {Bertaut}\ \emph {et~al.}(1961)\citenamefont
  {Bertaut}, \citenamefont {Corliss}, \citenamefont {Forrat}, \citenamefont
  {Aleonard},\ and\ \citenamefont {Pauthenet}}]{Bertaut1961}%
  \BibitemOpen
  \bibfield  {author} {\bibinfo {author} {\bibfnamefont {E.~F.}\ \bibnamefont
  {Bertaut}}, \bibinfo {author} {\bibfnamefont {L.}~\bibnamefont {Corliss}},
  \bibinfo {author} {\bibfnamefont {F.}~\bibnamefont {Forrat}}, \bibinfo
  {author} {\bibfnamefont {R.}~\bibnamefont {Aleonard}},\ and\ \bibinfo
  {author} {\bibfnamefont {R.}~\bibnamefont {Pauthenet}},\ }\bibfield  {title}
  {\bibinfo {title} {{Etude de niobates et tantalates de metaux de transition
  bivalents}},\ }\href
  {https://doi.org/https://doi.org/10.1016/0022-3697(61)90103-2} {\bibfield
  {journal} {\bibinfo  {journal} {Journal of Physics and Chemistry of Solids}\
  }\textbf {\bibinfo {volume} {21}},\ \bibinfo {pages} {234} (\bibinfo {year}
  {1961})}\BibitemShut {NoStop}%
\bibitem [{\citenamefont {Fang}\ \emph {et~al.}(2015)\citenamefont {Fang},
  \citenamefont {Zhou}, \citenamefont {Yan}, \citenamefont {Bai}, \citenamefont
  {Qian}, \citenamefont {Xu}, \citenamefont {Wang},\ and\ \citenamefont
  {Du}}]{Fang2015}%
  \BibitemOpen
  \bibfield  {author} {\bibinfo {author} {\bibfnamefont {Y.}~\bibnamefont
  {Fang}}, \bibinfo {author} {\bibfnamefont {W.~P.}\ \bibnamefont {Zhou}},
  \bibinfo {author} {\bibfnamefont {S.~M.}\ \bibnamefont {Yan}}, \bibinfo
  {author} {\bibfnamefont {R.}~\bibnamefont {Bai}}, \bibinfo {author}
  {\bibfnamefont {Z.~H.}\ \bibnamefont {Qian}}, \bibinfo {author}
  {\bibfnamefont {Q.~Y.}\ \bibnamefont {Xu}}, \bibinfo {author} {\bibfnamefont
  {D.~H.}\ \bibnamefont {Wang}},\ and\ \bibinfo {author} {\bibfnamefont
  {Y.~W.}\ \bibnamefont {Du}},\ }\bibfield  {title} {\bibinfo {title}
  {{Magnetic-field-induced dielectric anomaly and electric polarization in
  ${\mathrm{Mn}}_{4}{\mathrm{Nb}}_{2}{\mathrm{O}}_{9}$}},\ }\href
  {https://doi.org/10.1063/1.4913815} {\bibfield  {journal} {\bibinfo
  {journal} {Journal of Applied Physics}\ }\textbf {\bibinfo {volume} {117}},\
  \bibinfo {pages} {17B712} (\bibinfo {year} {2015})}\BibitemShut {NoStop}%
\bibitem [{\citenamefont {Panja}\ \emph {et~al.}(2021)\citenamefont {Panja},
  \citenamefont {Manuel},\ and\ \citenamefont {Nair}}]{Panja2021}%
  \BibitemOpen
  \bibfield  {author} {\bibinfo {author} {\bibfnamefont {S.~N.}\ \bibnamefont
  {Panja}}, \bibinfo {author} {\bibfnamefont {P.}~\bibnamefont {Manuel}},\ and\
  \bibinfo {author} {\bibfnamefont {S.}~\bibnamefont {Nair}},\ }\bibfield
  {title} {\bibinfo {title} {{Anisotropy in the magnetization and
  magnetoelectric response of single crystalline
  ${\mathrm{Mn}}_{4}{\mathrm{Ta}}_{2}{\mathrm{O}}_{9}$}},\ }\href
  {https://doi.org/10.1103/PhysRevB.103.014422} {\bibfield  {journal} {\bibinfo
   {journal} {Physical Review B}\ }\textbf {\bibinfo {volume} {103}},\ \bibinfo
  {pages} {014422} (\bibinfo {year} {2021})}\BibitemShut {NoStop}%
\bibitem [{\citenamefont {Khanh}\ \emph {et~al.}(2017)\citenamefont {Khanh},
  \citenamefont {Abe}, \citenamefont {Kimura}, \citenamefont {Tokunaga},\ and\
  \citenamefont {Arima}}]{Khanh2017}%
  \BibitemOpen
  \bibfield  {author} {\bibinfo {author} {\bibfnamefont {N.~D.}\ \bibnamefont
  {Khanh}}, \bibinfo {author} {\bibfnamefont {N.}~\bibnamefont {Abe}}, \bibinfo
  {author} {\bibfnamefont {S.}~\bibnamefont {Kimura}}, \bibinfo {author}
  {\bibfnamefont {Y.}~\bibnamefont {Tokunaga}},\ and\ \bibinfo {author}
  {\bibfnamefont {T.}~\bibnamefont {Arima}},\ }\bibfield  {title} {\bibinfo
  {title} {{Manipulation of electric polarization with rotating magnetic field
  in a honeycomb antiferromagnet
  ${\mathrm{Co}}_{\mathrm{4}}{\mathrm{Nb}}_{\mathrm{2}}{\mathrm{O}}_{\mathrm{9}}$}},\
  }\href {https://doi.org/10.1103/PhysRevB.96.094434} {\bibfield  {journal}
  {\bibinfo  {journal} {Physical Review B}\ }\textbf {\bibinfo {volume} {96}},\
  \bibinfo {pages} {094434} (\bibinfo {year} {2017})}\BibitemShut {NoStop}%
\bibitem [{\citenamefont {Lee}\ \emph {et~al.}(2020)\citenamefont {Lee},
  \citenamefont {Oh}, \citenamefont {Choi}, \citenamefont {Moon}, \citenamefont
  {Kim}, \citenamefont {Shin}, \citenamefont {Son}, \citenamefont {Nuss},
  \citenamefont {Kiryukhin},\ and\ \citenamefont {Choi}}]{Lee2020}%
  \BibitemOpen
  \bibfield  {author} {\bibinfo {author} {\bibfnamefont {N.}~\bibnamefont
  {Lee}}, \bibinfo {author} {\bibfnamefont {D.~G.}\ \bibnamefont {Oh}},
  \bibinfo {author} {\bibfnamefont {S.}~\bibnamefont {Choi}}, \bibinfo {author}
  {\bibfnamefont {J.~Y.}\ \bibnamefont {Moon}}, \bibinfo {author}
  {\bibfnamefont {J.~H.}\ \bibnamefont {Kim}}, \bibinfo {author} {\bibfnamefont
  {H.~J.}\ \bibnamefont {Shin}}, \bibinfo {author} {\bibfnamefont
  {K.}~\bibnamefont {Son}}, \bibinfo {author} {\bibfnamefont {J.}~\bibnamefont
  {Nuss}}, \bibinfo {author} {\bibfnamefont {V.}~\bibnamefont {Kiryukhin}},\
  and\ \bibinfo {author} {\bibfnamefont {Y.~J.}\ \bibnamefont {Choi}},\
  }\bibfield  {title} {\bibinfo {title} {{Highly nonlinear magnetoelectric
  effect in buckled-honeycomb antiferromagnetic
  ${\mathrm{Co}}_{4}{\mathrm{Ta}}_{2}{\mathrm{O}}_{9}$}},\ }\href
  {https://doi.org/https://doi.org/10.1038/s41598-020-69117-5} {\bibfield
  {journal} {\bibinfo  {journal} {Scientific Reports}\ }\textbf {\bibinfo
  {volume} {10}},\ \bibinfo {pages} {12362} (\bibinfo {year}
  {2020})}\BibitemShut {NoStop}%
\bibitem [{\citenamefont {Maignan}\ and\ \citenamefont
  {Martin}(2018{\natexlab{a}})}]{Maignan2018:2}%
  \BibitemOpen
  \bibfield  {author} {\bibinfo {author} {\bibfnamefont {A.}~\bibnamefont
  {Maignan}}\ and\ \bibinfo {author} {\bibfnamefont {C.}~\bibnamefont
  {Martin}},\ }\bibfield  {title} {\bibinfo {title} {{Type-II multiferroism and
  linear magnetoelectric coupling in the honeycomb
  $\mathrm{F}{\mathrm{e}}_{4}\mathrm{T}{\mathrm{a}}_{2}{\mathrm{O}}_{9}$
  antiferromagnet}},\ }\href
  {https://doi.org/10.1103/PhysRevMaterials.2.091401} {\bibfield  {journal}
  {\bibinfo  {journal} {Physical Review Materials}\ }\textbf {\bibinfo {volume}
  {2}},\ \bibinfo {pages} {091401} (\bibinfo {year}
  {2018}{\natexlab{a}})}\BibitemShut {NoStop}%
\bibitem [{\citenamefont {Zhang}\ \emph {et~al.}(2023)\citenamefont {Zhang},
  \citenamefont {Tang}, \citenamefont {Lin}, \citenamefont {Li}, \citenamefont
  {Zhou}, \citenamefont {Yang}, \citenamefont {Huang}, \citenamefont {Li},
  \citenamefont {Li}, \citenamefont {Zheng}, \citenamefont {Liu}, \citenamefont
  {Zeng}, \citenamefont {Wu}, \citenamefont {Yan}, \citenamefont {Huang},
  \citenamefont {Chen}, \citenamefont {Jiang},\ and\ \citenamefont
  {Liu}}]{Zhang2023}%
  \BibitemOpen
  \bibfield  {author} {\bibinfo {author} {\bibfnamefont {J.~H.}\ \bibnamefont
  {Zhang}}, \bibinfo {author} {\bibfnamefont {Y.~S.}\ \bibnamefont {Tang}},
  \bibinfo {author} {\bibfnamefont {L.}~\bibnamefont {Lin}}, \bibinfo {author}
  {\bibfnamefont {L.~Y.}\ \bibnamefont {Li}}, \bibinfo {author} {\bibfnamefont
  {G.~Z.}\ \bibnamefont {Zhou}}, \bibinfo {author} {\bibfnamefont
  {B.}~\bibnamefont {Yang}}, \bibinfo {author} {\bibfnamefont {L.}~\bibnamefont
  {Huang}}, \bibinfo {author} {\bibfnamefont {X.~Y.}\ \bibnamefont {Li}},
  \bibinfo {author} {\bibfnamefont {G.~Y.}\ \bibnamefont {Li}}, \bibinfo
  {author} {\bibfnamefont {S.~H.}\ \bibnamefont {Zheng}}, \bibinfo {author}
  {\bibfnamefont {M.~F.}\ \bibnamefont {Liu}}, \bibinfo {author} {\bibfnamefont
  {M.}~\bibnamefont {Zeng}}, \bibinfo {author} {\bibfnamefont {D.}~\bibnamefont
  {Wu}}, \bibinfo {author} {\bibfnamefont {Z.~B.}\ \bibnamefont {Yan}},
  \bibinfo {author} {\bibfnamefont {X.~K.}\ \bibnamefont {Huang}}, \bibinfo
  {author} {\bibfnamefont {C.}~\bibnamefont {Chen}}, \bibinfo {author}
  {\bibfnamefont {X.~P.}\ \bibnamefont {Jiang}},\ and\ \bibinfo {author}
  {\bibfnamefont {J.-M.}\ \bibnamefont {Liu}},\ }\bibfield  {title} {\bibinfo
  {title} {{Electric polarization reversal and nonlinear magnetoelectric
  coupling in the honeycomb antiferromagnet
  ${\mathrm{Fe}}_{4}{\mathrm{Nb}}_{2}{\mathrm{O}}_{9}$ single crystal}},\
  }\href {https://doi.org/10.1103/PhysRevB.107.024108} {\bibfield  {journal}
  {\bibinfo  {journal} {Physical Review B}\ }\textbf {\bibinfo {volume}
  {107}},\ \bibinfo {pages} {024108} (\bibinfo {year} {2023})}\BibitemShut
  {NoStop}%
\bibitem [{\citenamefont {Maignan}\ and\ \citenamefont
  {Martin}(2018{\natexlab{b}})}]{Maignan2018}%
  \BibitemOpen
  \bibfield  {author} {\bibinfo {author} {\bibfnamefont {A.}~\bibnamefont
  {Maignan}}\ and\ \bibinfo {author} {\bibfnamefont {C.}~\bibnamefont
  {Martin}},\ }\bibfield  {title} {\bibinfo {title}
  {{${\mathrm{Fe}}_{4}{\mathrm{Nb}}_{2}{\mathrm{O}}_{9}$: A magnetoelectric
  antiferromagnet}},\ }\href {https://doi.org/10.1103/PhysRevB.97.161106}
  {\bibfield  {journal} {\bibinfo  {journal} {Physical Review B}\ }\textbf
  {\bibinfo {volume} {97}},\ \bibinfo {pages} {161106} (\bibinfo {year}
  {2018}{\natexlab{b}})}\BibitemShut {NoStop}%
\bibitem [{\citenamefont {Ding}\ \emph {et~al.}(2020)\citenamefont {Ding},
  \citenamefont {Lee}, \citenamefont {Choi}, \citenamefont {Zhang},
  \citenamefont {Wu}, \citenamefont {Sinclair}, \citenamefont {Chakoumakos},
  \citenamefont {Chai}, \citenamefont {Zhou},\ and\ \citenamefont
  {Cao}}]{Ding2020}%
  \BibitemOpen
  \bibfield  {author} {\bibinfo {author} {\bibfnamefont {L.}~\bibnamefont
  {Ding}}, \bibinfo {author} {\bibfnamefont {M.}~\bibnamefont {Lee}}, \bibinfo
  {author} {\bibfnamefont {E.~S.}\ \bibnamefont {Choi}}, \bibinfo {author}
  {\bibfnamefont {J.}~\bibnamefont {Zhang}}, \bibinfo {author} {\bibfnamefont
  {Y.}~\bibnamefont {Wu}}, \bibinfo {author} {\bibfnamefont {R.}~\bibnamefont
  {Sinclair}}, \bibinfo {author} {\bibfnamefont {B.~C.}\ \bibnamefont
  {Chakoumakos}}, \bibinfo {author} {\bibfnamefont {Y.}~\bibnamefont {Chai}},
  \bibinfo {author} {\bibfnamefont {H.}~\bibnamefont {Zhou}},\ and\ \bibinfo
  {author} {\bibfnamefont {H.}~\bibnamefont {Cao}},\ }\bibfield  {title}
  {\bibinfo {title} {{Large spin-driven dielectric response and magnetoelectric
  coupling in the buckled honeycomb
  ${\mathrm{Fe}}_{4}{\mathrm{Nb}}_{2}{\mathrm{O}}_{9}$}},\ }\href
  {https://doi.org/10.1103/PhysRevMaterials.4.084403} {\bibfield  {journal}
  {\bibinfo  {journal} {Phys. Rev. Mater.}\ }\textbf {\bibinfo {volume} {4}},\
  \bibinfo {pages} {084403} (\bibinfo {year} {2020})}\BibitemShut {NoStop}%
\bibitem [{\citenamefont {Jana}\ \emph {et~al.}(2019)\citenamefont {Jana},
  \citenamefont {Sheptyakov}, \citenamefont {Ma}, \citenamefont {Alonso},
  \citenamefont {Pi}, \citenamefont {Mu\~noz}, \citenamefont {Liu},
  \citenamefont {Zhao}, \citenamefont {Su}, \citenamefont {Jin}, \citenamefont
  {Ma}, \citenamefont {Sun}, \citenamefont {Chen}, \citenamefont {Dong},
  \citenamefont {Chai}, \citenamefont {Li},\ and\ \citenamefont
  {Cheng}}]{Jana2019}%
  \BibitemOpen
  \bibfield  {author} {\bibinfo {author} {\bibfnamefont {R.}~\bibnamefont
  {Jana}}, \bibinfo {author} {\bibfnamefont {D.}~\bibnamefont {Sheptyakov}},
  \bibinfo {author} {\bibfnamefont {X.}~\bibnamefont {Ma}}, \bibinfo {author}
  {\bibfnamefont {J.~A.}\ \bibnamefont {Alonso}}, \bibinfo {author}
  {\bibfnamefont {M.}~\bibnamefont {Pi}}, \bibinfo {author} {\bibfnamefont
  {A.}~\bibnamefont {Mu\~noz}}, \bibinfo {author} {\bibfnamefont
  {Z.}~\bibnamefont {Liu}}, \bibinfo {author} {\bibfnamefont {L.}~\bibnamefont
  {Zhao}}, \bibinfo {author} {\bibfnamefont {N.}~\bibnamefont {Su}}, \bibinfo
  {author} {\bibfnamefont {S.}~\bibnamefont {Jin}}, \bibinfo {author}
  {\bibfnamefont {X.}~\bibnamefont {Ma}}, \bibinfo {author} {\bibfnamefont
  {K.}~\bibnamefont {Sun}}, \bibinfo {author} {\bibfnamefont {D.}~\bibnamefont
  {Chen}}, \bibinfo {author} {\bibfnamefont {S.}~\bibnamefont {Dong}}, \bibinfo
  {author} {\bibfnamefont {Y.}~\bibnamefont {Chai}}, \bibinfo {author}
  {\bibfnamefont {S.}~\bibnamefont {Li}},\ and\ \bibinfo {author}
  {\bibfnamefont {J.}~\bibnamefont {Cheng}},\ }\bibfield  {title} {\bibinfo
  {title} {{Low-temperature crystal and magnetic structures of the
  magnetoelectric material
  $\mathrm{F}{\mathrm{e}}_{4}\mathrm{N}{\mathrm{b}}_{2}{\mathrm{O}}_{9}$}},\
  }\href {https://doi.org/10.1103/PhysRevB.100.094109} {\bibfield  {journal}
  {\bibinfo  {journal} {Physical Review B}\ }\textbf {\bibinfo {volume}
  {100}},\ \bibinfo {pages} {094109} (\bibinfo {year} {2019})}\BibitemShut
  {NoStop}%
\bibitem [{\citenamefont {Choi}\ \emph {et~al.}(2020)\citenamefont {Choi},
  \citenamefont {Oh}, \citenamefont {Gutmann}, \citenamefont {Pan},
  \citenamefont {Kim}, \citenamefont {Son}, \citenamefont {Kim}, \citenamefont
  {Lee}, \citenamefont {Cheong}, \citenamefont {Choi},\ and\ \citenamefont
  {Kiryukhin}}]{Choi2020}%
  \BibitemOpen
  \bibfield  {author} {\bibinfo {author} {\bibfnamefont {S.}~\bibnamefont
  {Choi}}, \bibinfo {author} {\bibfnamefont {D.~G.}\ \bibnamefont {Oh}},
  \bibinfo {author} {\bibfnamefont {M.~J.}\ \bibnamefont {Gutmann}}, \bibinfo
  {author} {\bibfnamefont {S.}~\bibnamefont {Pan}}, \bibinfo {author}
  {\bibfnamefont {G.}~\bibnamefont {Kim}}, \bibinfo {author} {\bibfnamefont
  {K.}~\bibnamefont {Son}}, \bibinfo {author} {\bibfnamefont {J.}~\bibnamefont
  {Kim}}, \bibinfo {author} {\bibfnamefont {N.}~\bibnamefont {Lee}}, \bibinfo
  {author} {\bibfnamefont {S.-W.}\ \bibnamefont {Cheong}}, \bibinfo {author}
  {\bibfnamefont {Y.~J.}\ \bibnamefont {Choi}},\ and\ \bibinfo {author}
  {\bibfnamefont {V.}~\bibnamefont {Kiryukhin}},\ }\bibfield  {title} {\bibinfo
  {title} {{Noncollinear antiferromagnetic order in the buckled honeycomb
  lattice of magnetoelectric
  ${\mathrm{Co}}_{4}{\mathrm{Ta}}_{2}{\mathrm{O}}_{9}$ determined by
  single-crystal neutron diffraction}},\ }\href
  {https://doi.org/10.1103/PhysRevB.102.214404} {\bibfield  {journal} {\bibinfo
   {journal} {Physical Review B}\ }\textbf {\bibinfo {volume} {102}},\ \bibinfo
  {pages} {214404} (\bibinfo {year} {2020})}\BibitemShut {NoStop}%
\bibitem [{\citenamefont {Heid}\ \emph {et~al.}(1996)\citenamefont {Heid},
  \citenamefont {Weitzel}, \citenamefont {Bourdarot}, \citenamefont
  {Calemczuk}, \citenamefont {Vogt},\ and\ \citenamefont {Fuess}}]{Heid1996}%
  \BibitemOpen
  \bibfield  {author} {\bibinfo {author} {\bibfnamefont {C.}~\bibnamefont
  {Heid}}, \bibinfo {author} {\bibfnamefont {H.}~\bibnamefont {Weitzel}},
  \bibinfo {author} {\bibfnamefont {F.}~\bibnamefont {Bourdarot}}, \bibinfo
  {author} {\bibfnamefont {R.}~\bibnamefont {Calemczuk}}, \bibinfo {author}
  {\bibfnamefont {T.}~\bibnamefont {Vogt}},\ and\ \bibinfo {author}
  {\bibfnamefont {H.}~\bibnamefont {Fuess}},\ }\bibfield  {title} {\bibinfo
  {title} {{Magnetism in ${\mathrm{Fe}{\mathrm{Nb}}_{2}{\mathrm{O}}_6}$ and
  ${\mathrm{Ni}}{\mathrm{Nb}}_{2}{\mathrm{O}}_{6}$}},\ }\href
  {https://doi.org/10.1088/0953-8984/8/49/046} {\bibfield  {journal} {\bibinfo
  {journal} {Journal of Physics: Condensed Matter}\ }\textbf {\bibinfo {volume}
  {8}},\ \bibinfo {pages} {10609} (\bibinfo {year} {1996})}\BibitemShut
  {NoStop}%
\bibitem [{\citenamefont {Chapon}\ \emph {et~al.}(2011)\citenamefont {Chapon},
  \citenamefont {Manuel}, \citenamefont {Radaelli}, \citenamefont {Benson},
  \citenamefont {Perrott}, \citenamefont {Ansell}, \citenamefont {Rhodes},
  \citenamefont {Raspino}, \citenamefont {Duxbury}, \citenamefont {Spill},\
  and\ \citenamefont {Norris}}]{Chapon2011}%
  \BibitemOpen
  \bibfield  {author} {\bibinfo {author} {\bibfnamefont {L.~C.}\ \bibnamefont
  {Chapon}}, \bibinfo {author} {\bibfnamefont {P.}~\bibnamefont {Manuel}},
  \bibinfo {author} {\bibfnamefont {P.~G.}\ \bibnamefont {Radaelli}}, \bibinfo
  {author} {\bibfnamefont {C.}~\bibnamefont {Benson}}, \bibinfo {author}
  {\bibfnamefont {L.}~\bibnamefont {Perrott}}, \bibinfo {author} {\bibfnamefont
  {S.}~\bibnamefont {Ansell}}, \bibinfo {author} {\bibfnamefont {N.~J.}\
  \bibnamefont {Rhodes}}, \bibinfo {author} {\bibfnamefont {D.}~\bibnamefont
  {Raspino}}, \bibinfo {author} {\bibfnamefont {D.}~\bibnamefont {Duxbury}},
  \bibinfo {author} {\bibfnamefont {E.}~\bibnamefont {Spill}},\ and\ \bibinfo
  {author} {\bibfnamefont {J.}~\bibnamefont {Norris}},\ }\bibfield  {title}
  {\bibinfo {title} {Wish: {T}he new powder and single crystal magnetic
  diffractometer on the second target station},\ }\href
  {https://doi.org/10.1080/10448632.2011.569650} {\bibfield  {journal}
  {\bibinfo  {journal} {Neutron News}\ }\textbf {\bibinfo {volume} {22}},\
  \bibinfo {pages} {22} (\bibinfo {year} {2011})}\BibitemShut {NoStop}%
\bibitem [{\citenamefont {Aroyo}\ \emph {et~al.}(2011)\citenamefont {Aroyo},
  \citenamefont {Perez-Mato}, \citenamefont {Orobengoa}, \citenamefont {Tasci},
  \citenamefont {De~La~Flor},\ and\ \citenamefont {Kirov}}]{Aroyo2011}%
  \BibitemOpen
  \bibfield  {author} {\bibinfo {author} {\bibfnamefont {M.}~\bibnamefont
  {Aroyo}}, \bibinfo {author} {\bibfnamefont {J.}~\bibnamefont {Perez-Mato}},
  \bibinfo {author} {\bibfnamefont {D.}~\bibnamefont {Orobengoa}}, \bibinfo
  {author} {\bibfnamefont {E.}~\bibnamefont {Tasci}}, \bibinfo {author}
  {\bibfnamefont {G.}~\bibnamefont {De~La~Flor}},\ and\ \bibinfo {author}
  {\bibfnamefont {A.}~\bibnamefont {Kirov}},\ }\bibfield  {title} {\bibinfo
  {title} {{Crystallography online: Bilbao Crystallographic Server}},\ }\href
  {https://www.scopus.com/inward/record.uri?eid=2-s2.0-80955140447&partnerID=40&md5=488772b9e21d2636a3952f66ae80ae84}
  {\bibfield  {journal} {\bibinfo  {journal} {Bulgarian Chemical
  Communications}\ }\textbf {\bibinfo {volume} {43}},\ \bibinfo {pages} {183}
  (\bibinfo {year} {2011})}\BibitemShut {NoStop}%
\bibitem [{\citenamefont {Petříček}\ \emph {et~al.}(2014)\citenamefont
  {Petříček}, \citenamefont {Dušek},\ and\ \citenamefont
  {Palatinus}}]{Petricek2014}%
  \BibitemOpen
  \bibfield  {author} {\bibinfo {author} {\bibfnamefont {V.}~\bibnamefont
  {Petříček}}, \bibinfo {author} {\bibfnamefont {M.}~\bibnamefont
  {Dušek}},\ and\ \bibinfo {author} {\bibfnamefont {L.}~\bibnamefont
  {Palatinus}},\ }\bibfield  {title} {\bibinfo {title} {{Crystallographic
  Computing System {JANA}2006: General features}},\ }\href
  {https://doi.org/doi:10.1515/zkri-2014-1737} {\bibfield  {journal} {\bibinfo
  {journal} {Z. Kristallogr.}\ }\textbf {\bibinfo {volume} {229}},\ \bibinfo
  {pages} {345} (\bibinfo {year} {2014})}\BibitemShut {NoStop}%
\bibitem [{\citenamefont {Kresse}\ and\ \citenamefont
  {Furthm{\"u}ller}(1996)}]{Kresse1996}%
  \BibitemOpen
  \bibfield  {author} {\bibinfo {author} {\bibfnamefont {G.}~\bibnamefont
  {Kresse}}\ and\ \bibinfo {author} {\bibfnamefont {J.}~\bibnamefont
  {Furthm{\"u}ller}},\ }\bibfield  {title} {\bibinfo {title} {{Efficiency of
  ab-initio total energy calculations for metals and semiconductors using a
  plane-wave basis set}},\ }\href
  {https://doi.org/https://doi.org/10.1016/0927-0256(96)00008-0} {\bibfield
  {journal} {\bibinfo  {journal} {Computational Materials Science}\ }\textbf
  {\bibinfo {volume} {6}},\ \bibinfo {pages} {15} (\bibinfo {year}
  {1996})}\BibitemShut {NoStop}%
\bibitem [{\citenamefont {Perdew}\ \emph {et~al.}(1996)\citenamefont {Perdew},
  \citenamefont {Burke},\ and\ \citenamefont {Ernzerhof}}]{Perdew1996}%
  \BibitemOpen
  \bibfield  {author} {\bibinfo {author} {\bibfnamefont {J.~P.}\ \bibnamefont
  {Perdew}}, \bibinfo {author} {\bibfnamefont {K.}~\bibnamefont {Burke}},\ and\
  \bibinfo {author} {\bibfnamefont {M.}~\bibnamefont {Ernzerhof}},\ }\bibfield
  {title} {\bibinfo {title} {{Generalized Gradient Approximation Made
  Simple}},\ }\href {https://doi.org/10.1103/PhysRevLett.77.3865} {\bibfield
  {journal} {\bibinfo  {journal} {Physical Review Letters}\ }\textbf {\bibinfo
  {volume} {77}},\ \bibinfo {pages} {3865} (\bibinfo {year}
  {1996})}\BibitemShut {NoStop}%
\bibitem [{\citenamefont {Monkhorst}\ and\ \citenamefont
  {Pack}(1976)}]{Monkhorst1976}%
  \BibitemOpen
  \bibfield  {author} {\bibinfo {author} {\bibfnamefont {H.~J.}\ \bibnamefont
  {Monkhorst}}\ and\ \bibinfo {author} {\bibfnamefont {J.~D.}\ \bibnamefont
  {Pack}},\ }\bibfield  {title} {\bibinfo {title} {{Special points for
  Brillouin-zone integrations}},\ }\href
  {https://doi.org/10.1103/PhysRevB.13.5188} {\bibfield  {journal} {\bibinfo
  {journal} {Physical Review B}\ }\textbf {\bibinfo {volume} {13}},\ \bibinfo
  {pages} {5188} (\bibinfo {year} {1976})}\BibitemShut {NoStop}%
\bibitem [{\citenamefont {Bl\"ochl}(1994)}]{Blochl1994}%
  \BibitemOpen
  \bibfield  {author} {\bibinfo {author} {\bibfnamefont {P.~E.}\ \bibnamefont
  {Bl\"ochl}},\ }\bibfield  {title} {\bibinfo {title} {{Projector
  augmented-wave method}},\ }\href {https://doi.org/10.1103/PhysRevB.50.17953}
  {\bibfield  {journal} {\bibinfo  {journal} {Physical Review B}\ }\textbf
  {\bibinfo {volume} {50}},\ \bibinfo {pages} {17953} (\bibinfo {year}
  {1994})}\BibitemShut {NoStop}%
\bibitem [{\citenamefont {Kresse}\ and\ \citenamefont
  {Joubert}(1999)}]{Kresse1999}%
  \BibitemOpen
  \bibfield  {author} {\bibinfo {author} {\bibfnamefont {G.}~\bibnamefont
  {Kresse}}\ and\ \bibinfo {author} {\bibfnamefont {D.}~\bibnamefont
  {Joubert}},\ }\bibfield  {title} {\bibinfo {title} {{From ultrasoft
  pseudopotentials to the projector augmented-wave method}},\ }\href
  {https://doi.org/10.1103/PhysRevB.59.1758} {\bibfield  {journal} {\bibinfo
  {journal} {Physical Review B}\ }\textbf {\bibinfo {volume} {59}},\ \bibinfo
  {pages} {1758} (\bibinfo {year} {1999})}\BibitemShut {NoStop}%
\bibitem [{\citenamefont {Dudarev}\ \emph {et~al.}(1998)\citenamefont
  {Dudarev}, \citenamefont {Botton}, \citenamefont {Savrasov}, \citenamefont
  {Humphreys},\ and\ \citenamefont {Sutton}}]{Dudarev1998}%
  \BibitemOpen
  \bibfield  {author} {\bibinfo {author} {\bibfnamefont {S.~L.}\ \bibnamefont
  {Dudarev}}, \bibinfo {author} {\bibfnamefont {G.~A.}\ \bibnamefont {Botton}},
  \bibinfo {author} {\bibfnamefont {S.~Y.}\ \bibnamefont {Savrasov}}, \bibinfo
  {author} {\bibfnamefont {C.~J.}\ \bibnamefont {Humphreys}},\ and\ \bibinfo
  {author} {\bibfnamefont {A.~P.}\ \bibnamefont {Sutton}},\ }\bibfield  {title}
  {\bibinfo {title} {{Electron-energy-loss spectra and the structural stability
  of nickel oxide: An {LSDA+U} study}},\ }\href
  {https://doi.org/10.1103/PhysRevB.57.1505} {\bibfield  {journal} {\bibinfo
  {journal} {Phys. Rev. B}\ }\textbf {\bibinfo {volume} {57}},\ \bibinfo
  {pages} {1505} (\bibinfo {year} {1998})}\BibitemShut {NoStop}%
\bibitem [{\citenamefont {Ali}\ \emph {et~al.}(2024)\citenamefont {Ali},
  \citenamefont {Kim}, \citenamefont {Yadav}, \citenamefont {Lee},
  \citenamefont {Yoon},\ and\ \citenamefont {Choi}}]{Ali2024}%
  \BibitemOpen
  \bibfield  {author} {\bibinfo {author} {\bibfnamefont {A.}~\bibnamefont
  {Ali}}, \bibinfo {author} {\bibfnamefont {H.-S.}\ \bibnamefont {Kim}},
  \bibinfo {author} {\bibfnamefont {P.}~\bibnamefont {Yadav}}, \bibinfo
  {author} {\bibfnamefont {S.}~\bibnamefont {Lee}}, \bibinfo {author}
  {\bibfnamefont {D.}~\bibnamefont {Yoon}},\ and\ \bibinfo {author}
  {\bibfnamefont {S.}~\bibnamefont {Choi}},\ }\bibfield  {title} {\bibinfo
  {title} {{Partial molecular orbitals in face-sharing $3d$ manganese trimer:
  Comparative studies on
  ${\mathrm{Ba}}_{4}{\mathrm{TaMn}}_{3}{\mathrm{O}}_{12}$ and
  ${\mathrm{Ba}}_{4}{\mathrm{NbMn}}_{3}{\mathrm{O}}_{12}$}},\ }\href
  {https://doi.org/10.1103/PhysRevResearch.6.013231} {\bibfield  {journal}
  {\bibinfo  {journal} {Physical Review Research}\ }\textbf {\bibinfo {volume}
  {6}},\ \bibinfo {pages} {013231} (\bibinfo {year} {2024})}\BibitemShut
  {NoStop}%
\bibitem [{\citenamefont {Chen}\ \emph {et~al.}(2021)\citenamefont {Chen},
  \citenamefont {Luo}, \citenamefont {Ma}, \citenamefont {Xu}, \citenamefont
  {Kang}, \citenamefont {Lu}, \citenamefont {Zhang},\ and\ \citenamefont
  {Cao}}]{Chen2021}%
  \BibitemOpen
  \bibfield  {author} {\bibinfo {author} {\bibfnamefont {Y.}~\bibnamefont
  {Chen}}, \bibinfo {author} {\bibfnamefont {X.}~\bibnamefont {Luo}}, \bibinfo
  {author} {\bibfnamefont {X.}~\bibnamefont {Ma}}, \bibinfo {author}
  {\bibfnamefont {C.}~\bibnamefont {Xu}}, \bibinfo {author} {\bibfnamefont
  {B.}~\bibnamefont {Kang}}, \bibinfo {author} {\bibfnamefont {W.}~\bibnamefont
  {Lu}}, \bibinfo {author} {\bibfnamefont {J.}~\bibnamefont {Zhang}},\ and\
  \bibinfo {author} {\bibfnamefont {S.}~\bibnamefont {Cao}},\ }\bibfield
  {title} {\bibinfo {title} {{Zero-field ferroelectric state and
  magnetoelectric coupling in antiferromagnetic
  ${\mathrm{Fe}}_{4}{\mathrm{Nb}}_{2}{\mathrm{O}}_{9}$ single crystal}},\
  }\href {https://doi.org/https://doi.org/10.1016/j.ceramint.2020.12.028}
  {\bibfield  {journal} {\bibinfo  {journal} {Ceramics International}\ }\textbf
  {\bibinfo {volume} {47}},\ \bibinfo {pages} {9055} (\bibinfo {year}
  {2021})}\BibitemShut {NoStop}%
\bibitem [{\citenamefont {Chaudhary}\ \emph {et~al.}(2020)\citenamefont
  {Chaudhary}, \citenamefont {Nagpal},\ and\ \citenamefont
  {Patnaik}}]{Chaudhary2020}%
  \BibitemOpen
  \bibfield  {author} {\bibinfo {author} {\bibfnamefont {S.}~\bibnamefont
  {Chaudhary}}, \bibinfo {author} {\bibfnamefont {V.}~\bibnamefont {Nagpal}},\
  and\ \bibinfo {author} {\bibfnamefont {S.}~\bibnamefont {Patnaik}},\
  }\bibfield  {title} {\bibinfo {title} {{Magnetoelectric response in honeycomb
  antiferromagnet ${\mathrm{Fe}}_{4}{\mathrm{Nb}}_{2}{\mathrm{O}}_{9}$}},\
  }\href {https://doi.org/https://doi.org/10.1016/j.jmmm.2020.167305}
  {\bibfield  {journal} {\bibinfo  {journal} {Journal of Magnetism and Magnetic
  Materials}\ }\textbf {\bibinfo {volume} {515}},\ \bibinfo {pages} {167305}
  (\bibinfo {year} {2020})}\BibitemShut {NoStop}%
\bibitem [{\citenamefont {Singh}\ \emph {et~al.}(2023)\citenamefont {Singh},
  \citenamefont {Kaur}, \citenamefont {Kaur},\ and\ \citenamefont
  {Singh}}]{Singh2023}%
  \BibitemOpen
  \bibfield  {author} {\bibinfo {author} {\bibfnamefont {S.}~\bibnamefont
  {Singh}}, \bibinfo {author} {\bibfnamefont {A.}~\bibnamefont {Kaur}},
  \bibinfo {author} {\bibfnamefont {P.}~\bibnamefont {Kaur}},\ and\ \bibinfo
  {author} {\bibfnamefont {L.}~\bibnamefont {Singh}},\ }\bibfield  {title}
  {\bibinfo {title} {{High-temperature dielectric relaxation and electric
  conduction mechanisms in a
  ${\mathrm{La}}{\mathrm{Co}}{\mathrm{O}}_{3}$-modified
  ${\mathrm{Na}}_{0.5}{\mathrm{Bi}}_{0.5}{\mathrm{Ti}}_{3}{\mathrm{O}}_{3}$
  system}},\ }\href {https://doi.org/https://doi.org/10.1021/acsomega.3c04490}
  {\bibfield  {journal} {\bibinfo  {journal} {ACS omega}\ }\textbf {\bibinfo
  {volume} {8}},\ \bibinfo {pages} {25623} (\bibinfo {year}
  {2023})}\BibitemShut {NoStop}%
\bibitem [{\citenamefont {Zhang}\ \emph {et~al.}(2021)\citenamefont {Zhang},
  \citenamefont {Su}, \citenamefont {Mi}, \citenamefont {Pi}, \citenamefont
  {Zhou}, \citenamefont {Cheng},\ and\ \citenamefont {Chai}}]{Zhang2021}%
  \BibitemOpen
  \bibfield  {author} {\bibinfo {author} {\bibfnamefont {J.}~\bibnamefont
  {Zhang}}, \bibinfo {author} {\bibfnamefont {N.}~\bibnamefont {Su}}, \bibinfo
  {author} {\bibfnamefont {X.}~\bibnamefont {Mi}}, \bibinfo {author}
  {\bibfnamefont {M.}~\bibnamefont {Pi}}, \bibinfo {author} {\bibfnamefont
  {H.}~\bibnamefont {Zhou}}, \bibinfo {author} {\bibfnamefont {J.}~\bibnamefont
  {Cheng}},\ and\ \bibinfo {author} {\bibfnamefont {Y.}~\bibnamefont {Chai}},\
  }\bibfield  {title} {\bibinfo {title} {{Probing magnetic symmetry in
  antiferromagnetic ${\mathrm{Fe}}_{4}{\mathrm{Nb}}_{2}{\mathrm{O}}_{9}$ single
  crystals by linear magnetoelectric tensor}},\ }\href
  {https://doi.org/10.1103/PhysRevB.103.L140401} {\bibfield  {journal}
  {\bibinfo  {journal} {Physical Review B}\ }\textbf {\bibinfo {volume}
  {103}},\ \bibinfo {pages} {L140401} (\bibinfo {year} {2021})}\BibitemShut
  {NoStop}%
\bibitem [{\citenamefont {Willwater}\ \emph {et~al.}(2021)\citenamefont
  {Willwater}, \citenamefont {S\"ullow}, \citenamefont {Reehuis}, \citenamefont
  {Feyerherm}, \citenamefont {Amitsuka}, \citenamefont {Ouladdiaf},
  \citenamefont {Suard}, \citenamefont {Klicpera}, \citenamefont
  {Vali\ifmmode~\check{s}\else \v{s}\fi{}ka}, \citenamefont
  {Posp\'{\i}\ifmmode~\check{s}\else \v{s}\fi{}il},\ and\ \citenamefont
  {Sechovsk\'y}}]{Willwater2021}%
  \BibitemOpen
  \bibfield  {author} {\bibinfo {author} {\bibfnamefont {J.}~\bibnamefont
  {Willwater}}, \bibinfo {author} {\bibfnamefont {S.}~\bibnamefont {S\"ullow}},
  \bibinfo {author} {\bibfnamefont {M.}~\bibnamefont {Reehuis}}, \bibinfo
  {author} {\bibfnamefont {R.}~\bibnamefont {Feyerherm}}, \bibinfo {author}
  {\bibfnamefont {H.}~\bibnamefont {Amitsuka}}, \bibinfo {author}
  {\bibfnamefont {B.}~\bibnamefont {Ouladdiaf}}, \bibinfo {author}
  {\bibfnamefont {E.}~\bibnamefont {Suard}}, \bibinfo {author} {\bibfnamefont
  {M.}~\bibnamefont {Klicpera}}, \bibinfo {author} {\bibfnamefont
  {M.}~\bibnamefont {Vali\ifmmode~\check{s}\else \v{s}\fi{}ka}}, \bibinfo
  {author} {\bibfnamefont {J.}~\bibnamefont {Posp\'{\i}\ifmmode~\check{s}\else
  \v{s}\fi{}il}},\ and\ \bibinfo {author} {\bibfnamefont {V.}~\bibnamefont
  {Sechovsk\'y}},\ }\bibfield  {title} {\bibinfo {title} {Crystallographic and
  magnetic structure of {UNi}$_{4}$$^{11}${B}},\ }\href
  {https://doi.org/10.1103/PhysRevB.103.184426} {\bibfield  {journal} {\bibinfo
   {journal} {Phys. Rev. B}\ }\textbf {\bibinfo {volume} {103}},\ \bibinfo
  {pages} {184426} (\bibinfo {year} {2021})}\BibitemShut {NoStop}%
\bibitem [{\citenamefont {Yadav}\ \emph {et~al.}(2023)\citenamefont {Yadav},
  \citenamefont {Lee}, \citenamefont {Pascut}, \citenamefont {Kim},
  \citenamefont {Gutmann}, \citenamefont {Xu}, \citenamefont {Gao},
  \citenamefont {Cheong}, \citenamefont {Kiryukhin},\ and\ \citenamefont
  {Choi}}]{Yadav2023}%
  \BibitemOpen
  \bibfield  {author} {\bibinfo {author} {\bibfnamefont {P.}~\bibnamefont
  {Yadav}}, \bibinfo {author} {\bibfnamefont {S.}~\bibnamefont {Lee}}, \bibinfo
  {author} {\bibfnamefont {G.~L.}\ \bibnamefont {Pascut}}, \bibinfo {author}
  {\bibfnamefont {J.}~\bibnamefont {Kim}}, \bibinfo {author} {\bibfnamefont
  {M.~J.}\ \bibnamefont {Gutmann}}, \bibinfo {author} {\bibfnamefont
  {X.}~\bibnamefont {Xu}}, \bibinfo {author} {\bibfnamefont {B.}~\bibnamefont
  {Gao}}, \bibinfo {author} {\bibfnamefont {S.-W.}\ \bibnamefont {Cheong}},
  \bibinfo {author} {\bibfnamefont {V.}~\bibnamefont {Kiryukhin}},\ and\
  \bibinfo {author} {\bibfnamefont {S.}~\bibnamefont {Choi}},\ }\bibfield
  {title} {\bibinfo {title} {{Noncollinear magnetic order, in-plane anisotropy,
  and magnetoelectric coupling in the pyroelectric honeycomb antiferromagnet
  ${\mathrm{Ni}}_{2}{\mathrm{Mo}}_{3}{\mathrm{O}}_{8}$}},\ }\href
  {https://doi.org/10.1103/PhysRevResearch.5.033099} {\bibfield  {journal}
  {\bibinfo  {journal} {Physical Review Research}\ }\textbf {\bibinfo {volume}
  {5}},\ \bibinfo {pages} {033099} (\bibinfo {year} {2023})}\BibitemShut
  {NoStop}%
\bibitem [{\citenamefont {Katsura}\ \emph {et~al.}(2005)\citenamefont
  {Katsura}, \citenamefont {Nagaosa},\ and\ \citenamefont
  {Balatsky}}]{Katsura2005}%
  \BibitemOpen
  \bibfield  {author} {\bibinfo {author} {\bibfnamefont {H.}~\bibnamefont
  {Katsura}}, \bibinfo {author} {\bibfnamefont {N.}~\bibnamefont {Nagaosa}},\
  and\ \bibinfo {author} {\bibfnamefont {A.~V.}\ \bibnamefont {Balatsky}},\
  }\bibfield  {title} {\bibinfo {title} {{Spin Current and Magnetoelectric
  Effect in Noncollinear Magnets}},\ }\href
  {https://doi.org/10.1103/PhysRevLett.95.057205} {\bibfield  {journal}
  {\bibinfo  {journal} {Physical Review Letters}\ }\textbf {\bibinfo {volume}
  {95}},\ \bibinfo {pages} {057205} (\bibinfo {year} {2005})}\BibitemShut
  {NoStop}%
\bibitem [{\citenamefont {Sergienko}\ \emph {et~al.}(2006)\citenamefont
  {Sergienko}, \citenamefont {\ifmmode~\mbox{\c{S}}\else \c{S}\fi{}en},\ and\
  \citenamefont {Dagotto}}]{Sergienko2006}%
  \BibitemOpen
  \bibfield  {author} {\bibinfo {author} {\bibfnamefont {I.~A.}\ \bibnamefont
  {Sergienko}}, \bibinfo {author} {\bibfnamefont {C.}~\bibnamefont
  {\ifmmode~\mbox{\c{S}}\else \c{S}\fi{}en}},\ and\ \bibinfo {author}
  {\bibfnamefont {E.}~\bibnamefont {Dagotto}},\ }\bibfield  {title} {\bibinfo
  {title} {{Ferroelectricity in the Magnetic {\textit E}-Phase of Orthorhombic
  Perovskites}},\ }\href {https://doi.org/10.1103/PhysRevLett.97.227204}
  {\bibfield  {journal} {\bibinfo  {journal} {Physical Review Letters}\
  }\textbf {\bibinfo {volume} {97}},\ \bibinfo {pages} {227204} (\bibinfo
  {year} {2006})}\BibitemShut {NoStop}%
\bibitem [{\citenamefont {Arima}(2007)}]{Arima2007}%
  \BibitemOpen
  \bibfield  {author} {\bibinfo {author} {\bibfnamefont {T.-H.}\ \bibnamefont
  {Arima}},\ }\bibfield  {title} {\bibinfo {title} {{Ferroelectricity Induced
  by Proper-Screw Type Magnetic Order}},\ }\href
  {https://doi.org/10.1143/jpsj.76.073702} {\bibfield  {journal} {\bibinfo
  {journal} {Journal of the Physical Society of Japan}\ }\textbf {\bibinfo
  {volume} {76}},\ \bibinfo {pages} {073702} (\bibinfo {year}
  {2007})}\BibitemShut {NoStop}%
\bibitem [{\citenamefont {Choi}\ \emph {et~al.}(2021)\citenamefont {Choi},
  \citenamefont {Kiryukhin},\ and\ \citenamefont {Manuel}}]{Choi2019}%
  \BibitemOpen
  \bibfield  {author} {\bibinfo {author} {\bibfnamefont {S.}~\bibnamefont
  {Choi}}, \bibinfo {author} {\bibfnamefont {V.}~\bibnamefont {Kiryukhin}},\
  and\ \bibinfo {author} {\bibfnamefont {P.}~\bibnamefont {Manuel}},\
  }\bibfield  {title} {\bibinfo {title} {{Determination of successive magnetic
  orders of a new honeycomb-based multiferroic antiferromagnet}},\ }\bibfield
  {journal} {\bibinfo  {journal} {STFC ISIS Neutron and Muon Source}\ }\href
  {https://doi.org/10.5286/ISIS.E.RB1910501-1} {10.5286/ISIS.E.RB1910501-1}
  (\bibinfo {year} {2021})\BibitemShut {NoStop}%
\end{thebibliography}%

\end{document}